\documentclass[10pt,twocolumn,letterpaper]{article}

\usepackage[pagenumbers]{cvpr} 

\usepackage{booktabs}       
\usepackage{amsfonts}       
\usepackage{nicefrac}       
\usepackage{microtype}      
\usepackage{xcolor}         
\usepackage{amsmath}
\usepackage{subcaption}
\usepackage{makecell}
\usepackage{multirow}
\usepackage{soul}
\usepackage{enumitem}
\usepackage{float}
\usepackage{textcomp}

\newcommand{\myparagraph}[1]{{\vspace{.2em} \noindent \bf #1}}
\newcommand{\ieno}{\textit{i}.\textit{e}.}
\newcommand{\egno}{\textit{e}.\textit{g}.} 

\newcommand{\tcr}{\textcolor{red}}
\makeatletter
\renewcommand\@makefnmark{} 
\makeatother
\definecolor{cvprblue}{rgb}{0.21,0.49,0.74}
\usepackage[pagebackref,breaklinks,colorlinks=true, urlcolor=magenta]{hyperref}

\title{UniMIC: Towards Universal Multi-modality Perceptual Image Compression}

\author{
Yixin Gao, Xin Li\textsuperscript{*}, Xiaohan Pan, Runsen Feng, Zongyu Guo, Yiting Lu, Yulin Ren, Zhibo Chen\thanks{Xin Li and Zhibo Chen are corresponding authors.} \\
University of Science and Technology of China \\
\tt\small \{gaoyixin, pxh123, fengruns, guozy, luyt31415, renyulin\}@mail.ustc.edu.cn, \\
\tt\small \{xin.li, chenzhibo\}@ustc.edu.cn
}

\begin{document}
\maketitle
\begin{abstract}
We present UniMIC, a universal multi-modality image compression framework, intending to unify the rate-distortion-perception (RDP) optimization for multiple image codecs simultaneously through excavating cross-modality generative priors. Unlike most existing works that need to design and optimize image codecs from scratch, our UniMIC introduces the visual codec repository, which incorporates amounts of representative image codecs and directly uses them as the basic codecs for various practical applications. Moreover, we propose multi-grained textual coding, where variable-length content prompt and compression prompt are designed and encoded to assist the perceptual reconstruction through the multi-modality conditional generation. 
In particular, a universal perception compensator is proposed to improve the perception quality of decoded images from all basic codecs at the decoder side by reusing text-assisted diffusion priors from stable diffusion. With the cooperation of the above three strategies, our UniMIC achieves a significant improvement of RDP optimization for different compression codecs, \egno, traditional and learnable codecs, and different compression costs, \egno, ultra-low bitrates. The code will be available in \href{https://github.com/Amygyx/UniMIC}{https://github.com/Amygyx/UniMIC}.
\end{abstract}
\section{Introduction}
\label{sec:intro}

\begin{figure*}[tb]
    \centering
    \captionsetup[subfigure]{labelformat=empty,font=tiny, aboveskip=0pt, belowskip=0pt, justification=centering}
    \begin{subfigure}[b]{0.245\textwidth}
        \includegraphics[width=\textwidth]{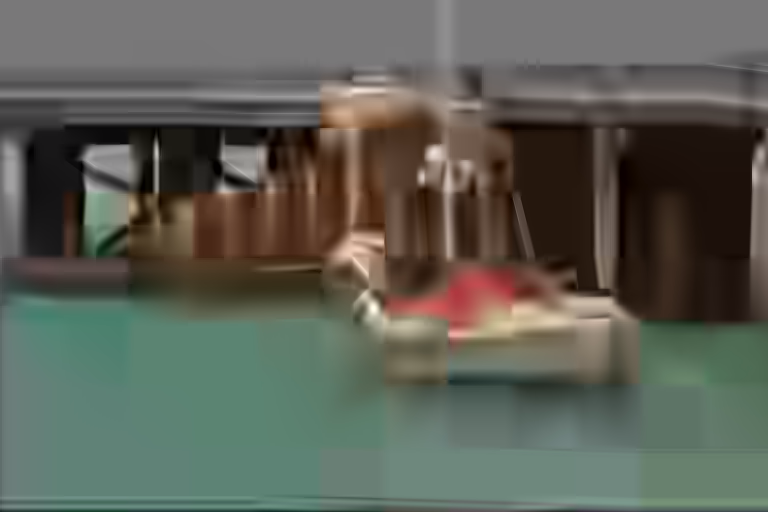}
        \caption{VTM, 0.0079bpp}
    \end{subfigure}
    \begin{subfigure}[b]{0.245\textwidth}
        \includegraphics[width=\textwidth]{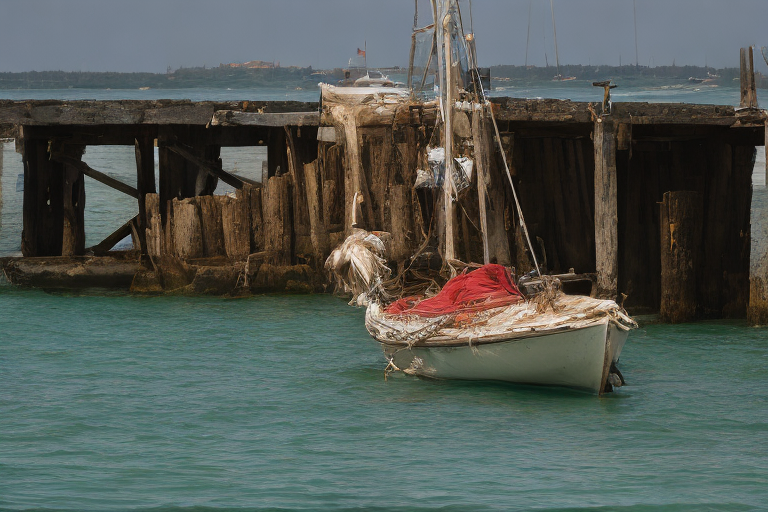}
        \caption{UniMIC(VTM-based), 0.0134bpp}
    \end{subfigure}
    \begin{subfigure}[b]{0.245\textwidth}
        \includegraphics[width=\textwidth]{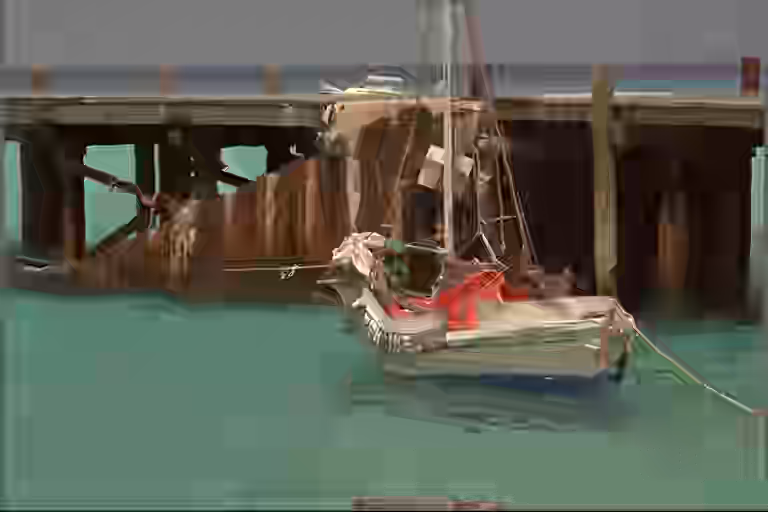}
        \caption{HM, 0.0437bpp}
    \end{subfigure}
    \begin{subfigure}[b]{0.245\textwidth}
        \includegraphics[width=\textwidth]{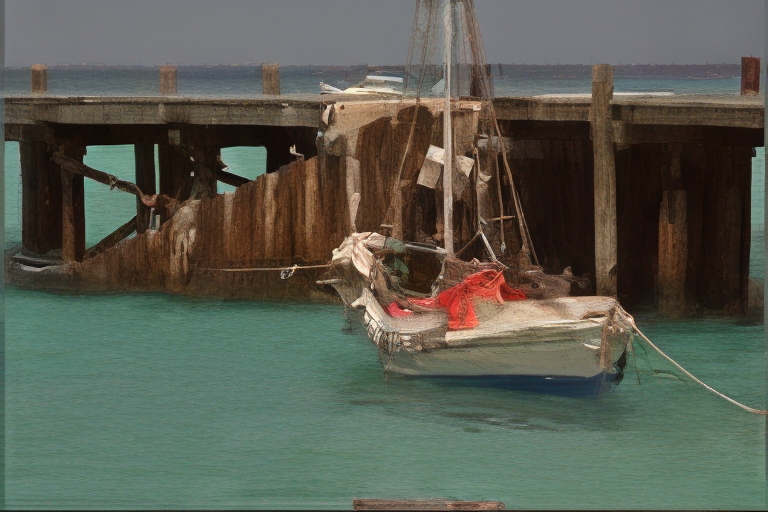}
        \caption{UniMIC(HM-based), 0.0492bpp}
    \end{subfigure}
    \begin{subfigure}[b]{0.245\textwidth}
        \includegraphics[width=\textwidth]{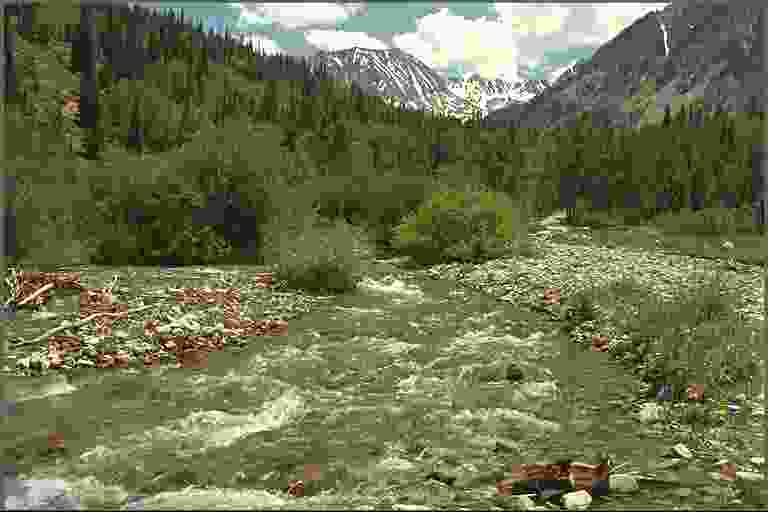}
        \caption{JPEG, 0.2945bpp}
    \end{subfigure}
    \begin{subfigure}[b]{0.245\textwidth}
        \includegraphics[width=\textwidth]{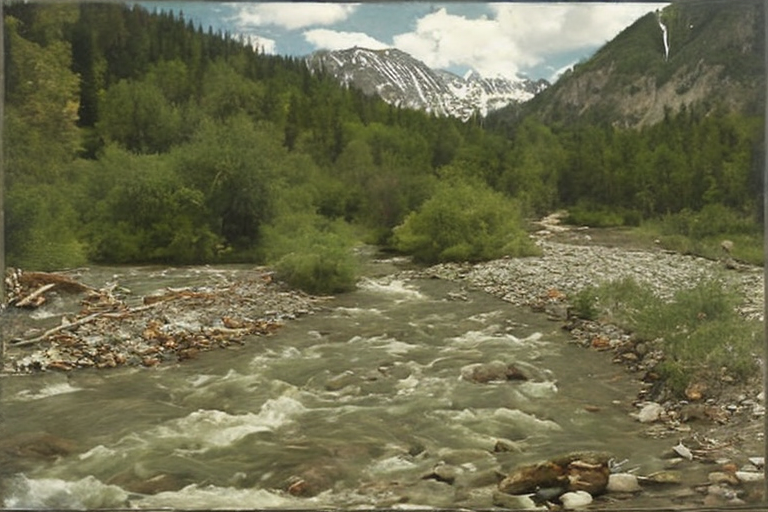}
        \caption{UniMIC (JPEG-based), 0.2986bpp}
    \end{subfigure}
    \begin{subfigure}[b]{0.245\textwidth}
        \includegraphics[width=\textwidth]{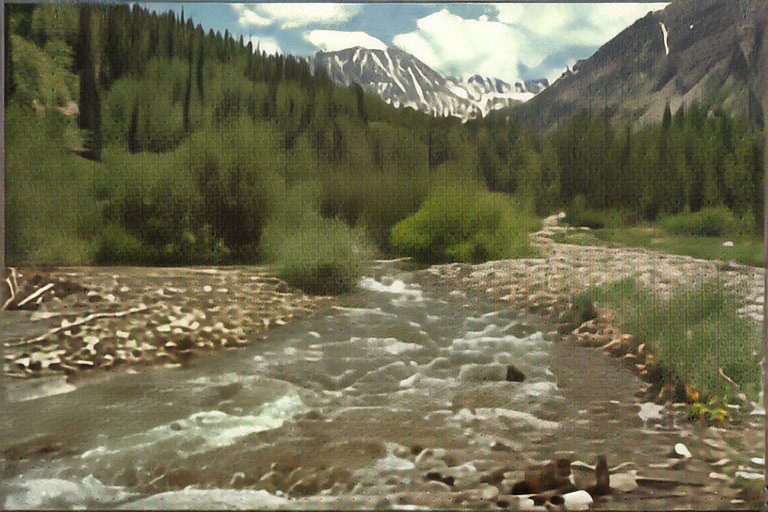}
        \caption{MS-ILLM, 0.0581bpp}
    \end{subfigure}
    \begin{subfigure}[b]{0.245\textwidth}
        \includegraphics[width=\textwidth]{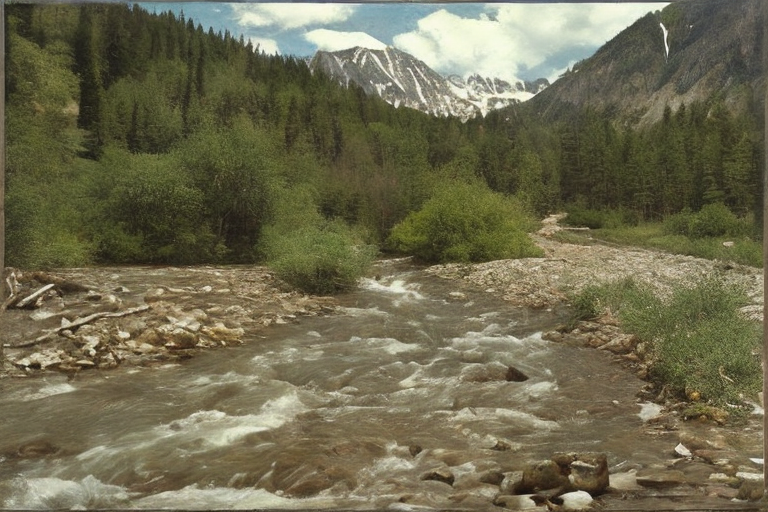}
        \caption{UniMIC (MS-ILLM-based), 0.0622bpp}
    \end{subfigure}
    \begin{subfigure}[b]{0.245\textwidth}
        \includegraphics[width=\textwidth]{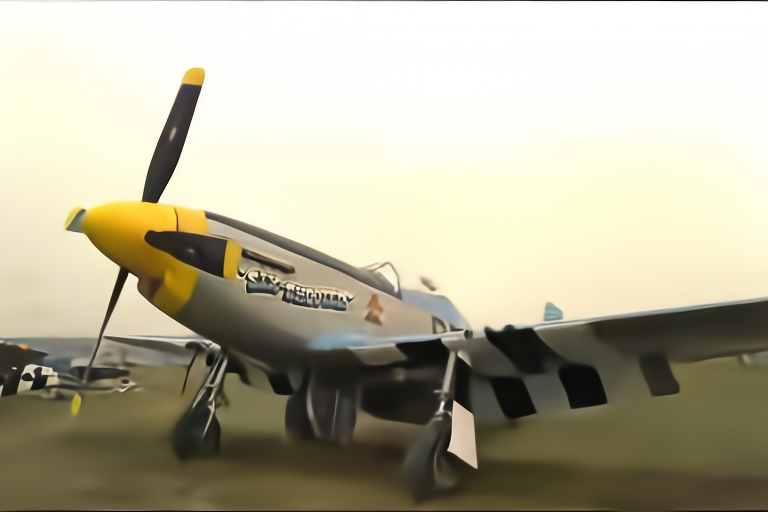}
        \caption{ELIC, 0.0279bpp}
    \end{subfigure}
    \begin{subfigure}[b]{0.245\textwidth}
        \includegraphics[width=\textwidth]{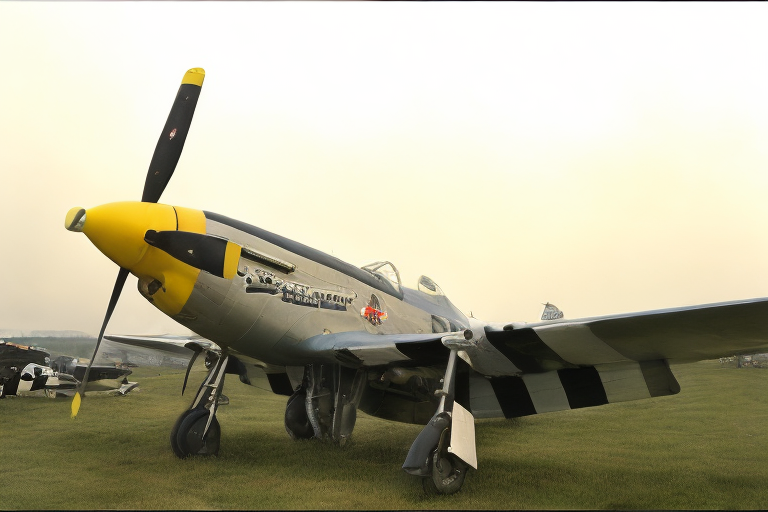}
        \caption{UniMIC (ELIC-based), 0.0319bpp}
    \end{subfigure}
    \begin{subfigure}[b]{0.245\textwidth}
        \includegraphics[width=\textwidth]{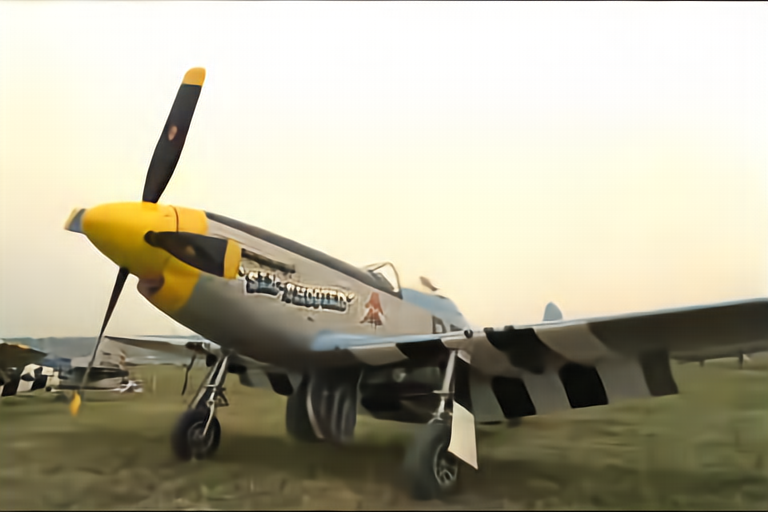}
        \caption{mbt2018, 0.0729bpp}
    \end{subfigure}
    \begin{subfigure}[b]{0.245\textwidth}
        \includegraphics[width=\textwidth]{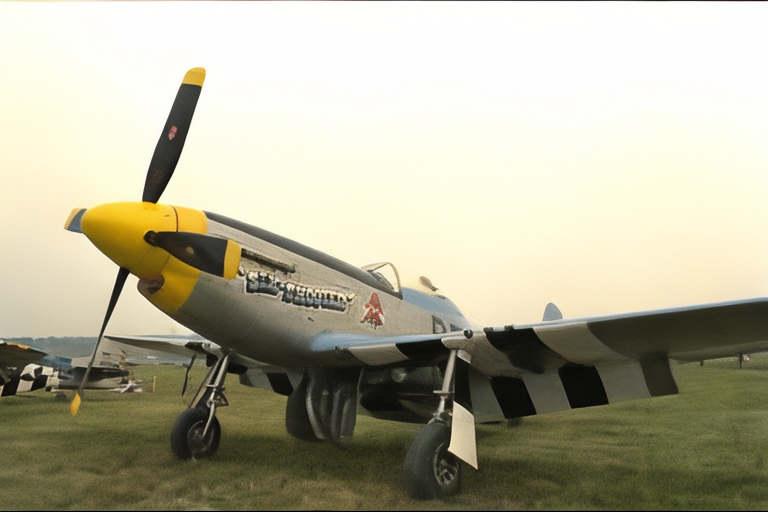}
        \caption{UniMIC (mbt2018-based), 0.0769bpp}
    \end{subfigure}
    
    \begin{subfigure}[b]{0.245\textwidth}
        \includegraphics[width=\textwidth]{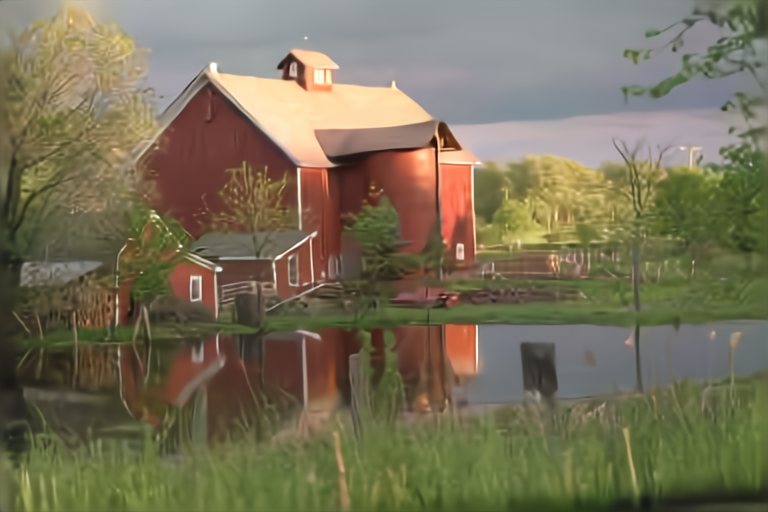}
        \caption{cheng20-msssim, \\ 0.1138bpp}
    \end{subfigure}
    \begin{subfigure}[b]{0.245\textwidth}
        \includegraphics[width=\textwidth]{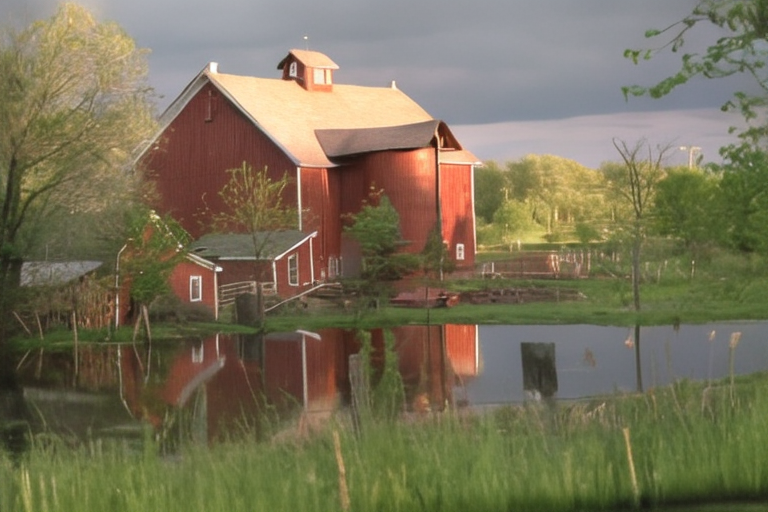}
        \caption{UniMIC (cheng20-msssim-based),\\ 0.1184bpp}
    \end{subfigure}
    \begin{subfigure}[b]{0.245\textwidth}
        \includegraphics[width=\textwidth]{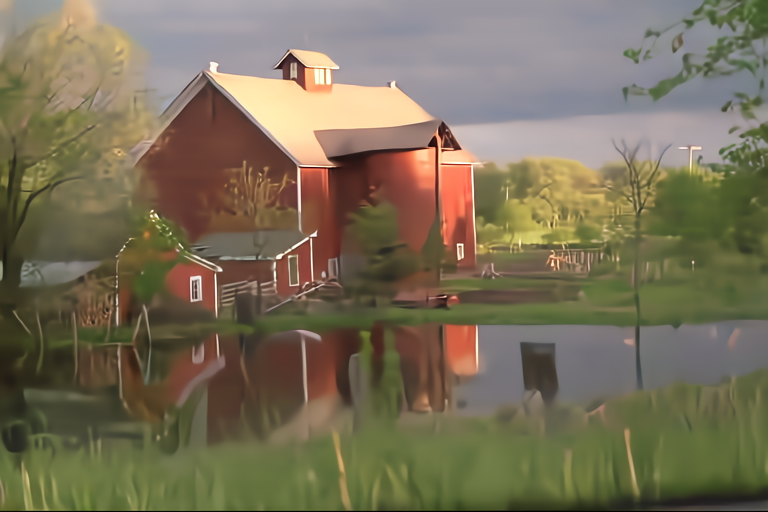}
        \caption{cheng20-mse,\\0.0901bpp}
    \end{subfigure}
    \begin{subfigure}[b]{0.245\textwidth}
        \includegraphics[width=\textwidth]{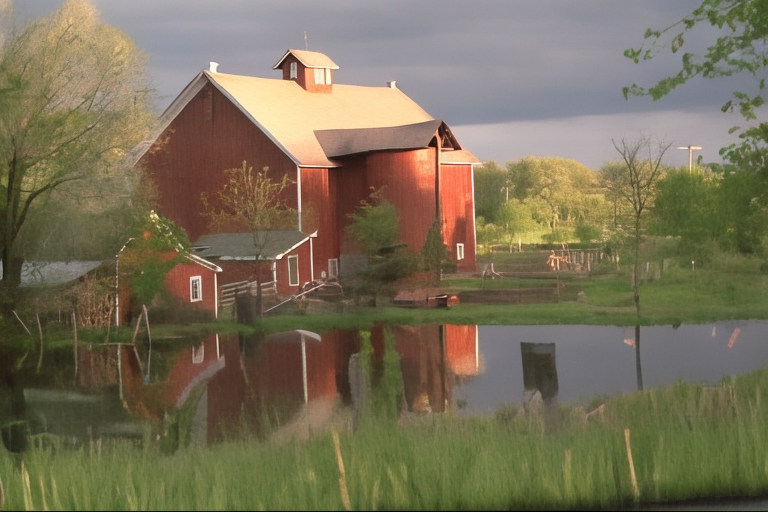}
        \caption{UniMIC (cheng20-mse-based),\\0.0947bpp}
    \end{subfigure}
    \caption{\small Visual comparisons of our proposed UniMIC framework with eight representative basic codecs, including the hand-crafted codec VTM~\cite{VVC}, HM~\cite{HEVC}, JPEG\cite{wallace1992jpeg}, MSE-optimized neural codec ELIC~\cite{he2022elic}, cheng20-mse~\cite{cheng2020learned}, mbt2018~\cite{minnen2018joint}, MS-SSIM-optimized cheng20-msssim on Kodak Dateset. Our method achieves more realistic and clear reconstructions than all basic codecs.}
    \label{fig:concept}
\end{figure*}
Recent years have witnessed the great advancement of image compression~\cite{liu2023learned, he2022elic,guo2021causal,wu2021learned,cheng2020learned,minnen2020channel,minnen2018joint,li2021task} in improving the rate-distortion optimization when compressing the images into the bit streams. Early traditional image codecs, \egno, JPEG\cite{wallace1992jpeg} and BPG\cite{bellard_bpg}, remove the content redundancy through several hand-crafted techniques, \egno, intra-frame prediction, transform coding and entropy coding. In contrast, neural image compression~\cite{cheng2020learned,zhu2022transformer,yang2020improving,guo2021soft,feng2023nvtc,balle2018variational,minnen2018joint,minnen2020channel,he2022elic} exploits deep neural networks to establish the nonlinear transform, quantization, and entropy modeling in an end-to-end training manner, thereby owning the potential to model the data distribution more accurately and adaptively than traditional codecs. Based on the optimization objectives, existing works on image compression can be roughly divided into three types: (i) improving the objective quality of decoded images with the same bitrate~\cite{liu2023learned, he2022elic,guo2021causal,cheng2020learned,minnen2020channel,minnen2018joint}, \egno, PSNR and MS-SSIM; (ii) enhancing the perceptual quality~\cite{agustsson2019generative,mentzer2020high,msillm,yang2024lossy,xu2024idempotence}; and (iii) assisting in the application of machine vision~\cite{sun2020semantic,feng2023semantically}. However, most works are still designed based on the single-modality compression, \ieno, only reducing the redundancy in visual space, which overlooked the cross-modality redundancy modeling and information compensation. 



Thanks to the fast development of cross-modality generative foundation models, \egno, Stable Diffusion~\cite{rombach2022high} and DALLE-3~\cite{dalle3}, some works have taken the first step to investigate how to incorporate the textual modality as the assistance for perceptual image compression, \ieno, enhancing the perceptual quality of decoded images, constructing the multi-modality compression. For instance, Lei \etal\cite{lei2023text} propose compressing images with text and sketch based on a conditional diffusion model. PerCo~\cite{perco} achieves realistic reconstructions even at 0.003 bpp with vector-quantized latent and a concise caption through fine-tuning a large-scale text-to-image diffusion model. Compared with existing GAN/Diffusion-based perceptual image compression, multi-modality compression enables ultra low-bitrate compression~\cite{perco} and better perceptual quality. 
Nonetheless, the above works entail the redesign of codecs for different coding requirements/scenarios and lack the universality for perceptual image compression at different bitrates.

In this work, we propose the first universal multi-modality perceptual image compression framework, dubbed UniMIC, aiming at establishing unified rate-distortion-perception (RDP) optimization by excavating the cross-modality generation priors for multiple codecs simultaneously. Instead of incorporating the textual modality by designing new image codecs like existing works~\cite{taco,qin2023perceptual}, which can merely support limited applications, \egno, the ultra low-bitrate compression, our UniMIC can reuse the visual compression capability of various popular and mature image codecs by constructing the compositional codec repository. The compositional codec repository is composed of traditional codecs, \egno, JPEG~\cite{wallace1992jpeg}, HEVC~\cite{HEVC} and VVC~\cite{VVC}, learned-based codecs, including ELIC~\cite{he2022elic}, and MS-ILLM~\cite{msillm}, etc, covering broad application scenarios in both industry and academic fields. To achieve universal textual modality compression suitable for all these basic visual codecs, we propose multi-grained textual coding as the information compensation from textual modality. In particular, three-level variable-length content prompt coding modes are provided to transfer the content descriptions of images with different lengths to the decoding side, where a longer content description will deliver more visual information about images. It is noteworthy that our universal perceptual image compression entails the adaptation for various compression codecs. Consequently, we also design the compression prompt as part of textual coding, which intends to transfer the compression syntax, including the type of basic codec, compression quality, and optimization objective, as the guidance on the decoding side. 

To achieve universal perceptual optimization with textual modality, we propose the universal perception compensator by excavating the cross-modality generation priors from pre-trained large generative foundation models, \ieno, Stable Diffusion~\cite{rombach2022high}. Concretely, we design the universal visual adapter, which incorporates the compression prompts, \ieno, compression syntax, and decoded visual images into the perceptual compensator through scale and shift modification, thereby enabling the adaptation capability for different codecs. The content prompt is injected into the perceptual compensator by reusing the cross-modality generation module in stable diffusion.  
Based on the above three innovations, our UniMic significantly improves the RDP optimization and achieves multi-modality compression for all image codecs in our proposed compositional codec repository at different bitrate points, and owns great applicability to unseen compression codecs.  

The main contributions of our approach are summarized as follows:
\begin{itemize}[left=0pt]
\item We propose the first universal multi-modality perceptual image compression framework, which establishes the unified rate-distortion-perception (RDP) optimization for all image codecs of our proposed compositional codec repository.


\item To achieve universal multi-modality compression, we propose multi-grained textual coding, where three-level variable-length content prompt coding can assist the requirements on different bitrates, and transfer the content information to the decoding side for textual-assisted decoding. A compression prompt coding is designed to achieve the adaptability and applicability of UniMIC for different basic codecs. 
\item To achieve the RDP optimization, we design a universal perception compensator, which reuses the cross-modality generation priors of diffusion models by designing the universal visual adapter under the guidance of transferred textual information.
\end{itemize}

\section{Related Works}
\label{sec:related_works}

\subsection{Traditional Image Compression}
Traditional image codecs, such as JPEG~\cite{wallace1992jpeg}, JPEG2000 \cite{rabbani2002overview}, BPG \cite{bellard_bpg}, HEVC \cite{HEVC}, VVC \cite{VVC}, have been developed over several decades and are widely used today.
These standards rely on transform coding~\cite{goyal2001theoretical}, which mainly consists of three components: transform, quantization, and entropy coding. Each module in these standards is manually designed with multiple modes, and rate-distortion optimization is performed to select the optimal mode. However, because these modules are hand-crafted, traditional image codecs are limited in their ability to adapt to various objectives, \egno, perceptual optimization and classification accuracy, as they cannot be optimized in an end-to-end manner.

\subsection{Generative Image Compression} 
Blau \etal~\cite{blau2018perception,blau2019rethinking} formalized the rate-distortion-perception (RDP) tradeoff, defining perceptual quality as the divergence between real and reconstructed image distributions. 
A series of works~\cite{agustsson2019generative, mentzer2020high, msillm} incorporate adversarial loss~\cite{goodfellow2014generative} to reduce the mismatch between the compressed and natural image distributions. Recently, given the great success of diffusion models in image generation, some recent works explore combining diffusion models for better perceptual quality. 
{ These methods can be roughly divided into two categories, 1) end-to-end RDP optimization. Theis \etal~\cite{theis2022lossy} develop an approach that integrates diffusion model with reverse channel coding techniques~\cite{havasi2019minimal, theis2022algorithms}, demonstrating that it outperforms HiFiC~\cite{mentzer2020high} on 64 $\times$ 64 ImageNet images. CDC~\cite{yang2024lossy} employs a VAE-style encoder to map images to latent variables and a diffusion model as the decoder, conditioned on quantized latents.
2) The second type is refining images based on existing neural image codecs~\cite{ghouse2023residual,hoogeboom2023high,xu2024idempotence}. For example, HFD~\cite{hoogeboom2023high} uses a pretrained neural codec and refines its output with a diffusion model. 
However, these approaches rely on single-modality, focusing solely on reducing visual information redundancy. In contrast, our model leverages cross-modality redundancy by utilizing multi-grained text information to enhance perceptual reconstruction.}

\subsection{Neural Image Compression with Multi-modality Information}


Multi-modality information (\egno, auditory, textual, LiDAR) is widely used in various computer vision tasks~\cite{uppal2022multimodal,acosta2022multimodal,huang2022multi} for high accuracy or robust perception. 
A series of works leverage multi-modality information for better semantic similarity and human-comprehensible communication. Some studies ~\cite{chang2021thousand,CMC,rdocmc,lei2023text} represent images solely by other 
modalities (\egno, text, edge map, semantic map) for semantic image compression at ultra-low bitrates. 
Furthermore, some methods~\cite{jiang2023multi,perco,qin2023perceptual,taco,pan2022extreme,li2024misc} utilize multi-modality information (especially text description of semantic information) to guide the transmission and reconstruction of compressed image representations. More specifically, most of these methods~\cite{jiang2023multi,perco,qin2023perceptual,taco} introduce a brief image caption to supplement high-level semantic information. 
While Li \etal\cite{li2024misc} propose adding object-level descriptions and performing multiple text-guided image restoration processes on a downsampled and compressed image, resulting in high decoding complexity. 
Nonetheless, these approaches require the redesign of codecs for different coding requirements or scenarios and lack the universality for perceptual image compression at various bitrates. Differently, our approach not only uses text to describe semantic information but also employs it to convey compression-related information, enhancing adaptability across different codecs and bitrates.
\section{Methods}
\label{sec:methods}

\begin{figure*}[tb]
\centerline{\includegraphics[width=1.0\linewidth]{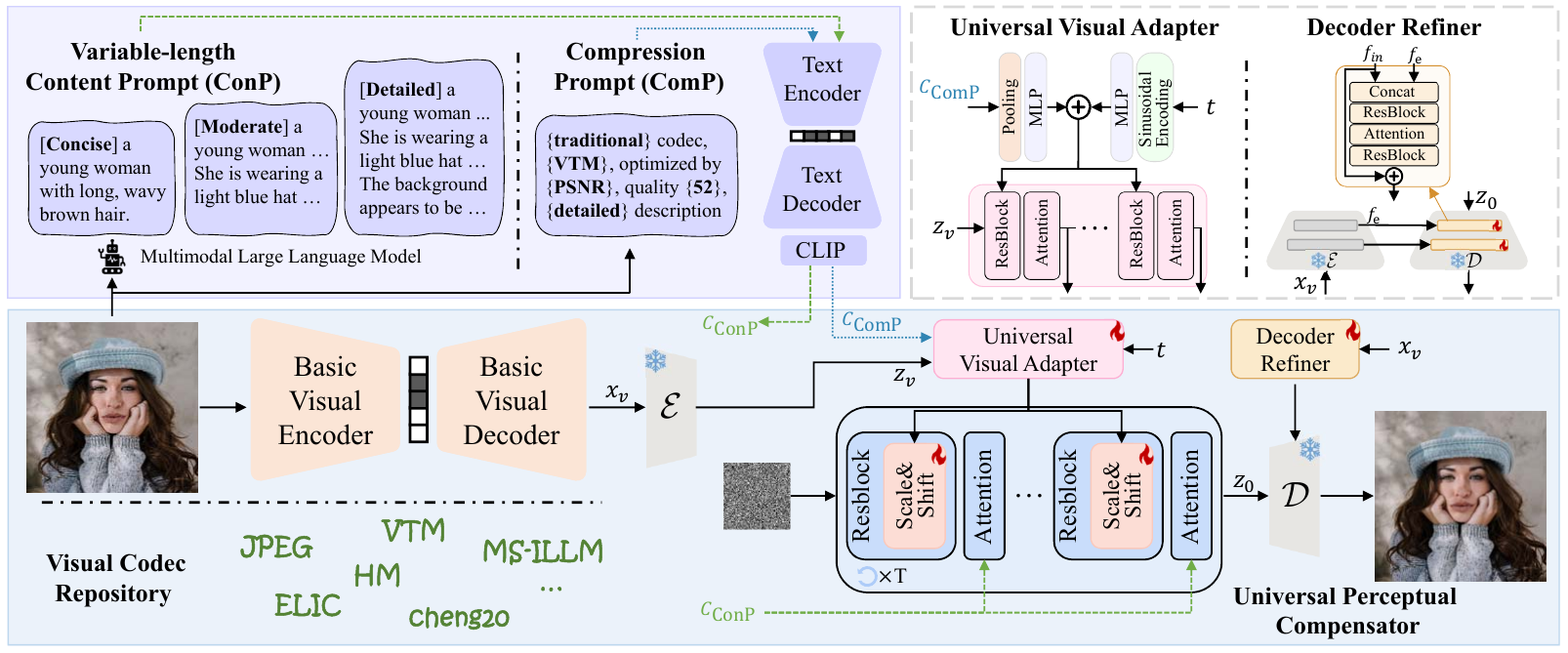}}
    \caption{Illustration of our proposed UniMIC. The visual codec repository includes various representative basic codecs and provides visual representations. Multi-grained textual information, composed of variable-length content prompt and compression prompt, is losslessly transmitted and processed by CLIP text encoder. Finally, the universal perceptual compensator takes the multi-modality information to conduct the diffusion process.}
\label{fig:overview}
\end{figure*}
\begin{table*}[h]
    \centering
    \begin{tabular}{cccc}
    \toprule
    Level &  Length & Description Aspect & Average Bytes on DIV2K\\
    \midrule
    Concise&  $\leq$16 words & Overall & 67.01 \\
    Moderate& $\leq$36 words & + foreground & 122.53 \\
    Detailed& $\leq$75 words & \begin{tabular}[c]{@{}c@{}} + more detailed specifics and background \end{tabular} & 239.43 \\
    \bottomrule
    \end{tabular}
    \vspace{-2mm}
    \caption{Description of our variable-length content prompt.}
    \label{tab:illustration_of_ConP}
\end{table*}

\subsection{Preliminary}
\myparagraph{Diffusion models}~\cite{sohl2015deep, ho2020denoising, song2020score,li2023diffusion} are probabilistic models designed to learn a data distribution by gradual denoising a normally distributed variable. Our work is based on Stable Diffusion~\cite{rombach2022high}, which conducts diffusion process in latent space, facilitating high-resolution synthesis and computational efficiency. The model comprises two components: an autoencoder ($\mathcal{E}(\cdot), \mathcal{D}(\cdot)$) and an UNet denoiser $\epsilon_{\theta}$, and employs a two-stage training process. First, the autoencoder is trained to achieve $\mathcal{D}(\mathcal{E}(x)) \approx x$. Then, with the encoded latent of an image $z = \mathcal{E}(x)$, the optimization objective of UNet denoiser is 
\begin{equation}
    \mathbb{E}_{z \sim \mathcal{E}(x), t, C, \epsilon \sim \mathcal{N}(0,1)} \left[ \left\| \epsilon - \epsilon_{\theta} \left( z_t, t, C) \right) \right\|^2 \right],
\label{eq: SD_train}
\end{equation}
where $\epsilon$ represents the unscaled noise, $t$ is the time step, $z_t$ is the latent noise at time $t$, and $C$ denotes the textual embedding of prompts extracted by a pre-trained CLIP text encoder. During inference, a random Gaussian noise $z_T$ is iteratively denoised to $z_0$, and the final image is obtained through the decoder $x' = \mathcal{D}(z_0)$.

\myparagraph{Classifier-free guidance}~\cite{ho2021classifierfree} is a widely-used technique to improve sample quality while reducing diversity in conditional diffusion models through jointly training a single diffusion model on conditional and unconditional objectives via randomly dropping $C$ during training (e.g. with 10\% probability). Sampling is performed as
\begin{equation}
\tilde{\epsilon}_{\theta}(z_t, t, C) = w \epsilon_{\theta}(z_t, t, C) + (1 - w) \epsilon_{\theta}(z_t, t, \emptyset).
\label{eq: Classifier_sample}
\end{equation}
Here, $\epsilon_{\theta}(z_t, t, C)$ and $\epsilon_{\theta}(z_t, t, \emptyset)$ are conditional and unconditional $\epsilon$-predictions, and $w$ is the guidance weight. Setting $w = 1$ disables classifier-free guidance, while increasing $w > 1$ strengthens the effect of guidance. Stable Diffusion depends critically on classifier-free guidance for effective text conditioning.

\subsection{Overview}
The overall framework structure is illustrated in Fig.~\ref{fig:overview}. We present a universal multi-modality image compression framework, dubbed UniMIC, which unifies the rate-distortion-perception optimization for various image codecs with multi-modality information.
To achieve this, we first build a visual codec repository, which is composed of multiple representative codecs from several families (\egno, MSE-optimized codecs, including both traditional and neural, MS-SSIM optimized, and GAN loss-optimized) to simulate the diversity of image codec deployments in real-world scenarios as much as possible. Based on this, we design multi-grained textual coding and a universal perceptual compensator to improve the perception quality of decoded images for all basic codecs based on the powerful cross-modality generation prior of Stable Diffusion~\cite{rombach2022high}, which will be introduced in detail in the following sections. 

\subsection{Multi-grained Textual Coding}
\label{sec:Methods_textual_coding}
\myparagraph{Variable-length content prompt.} Motivated by the successful application of synthetic highly detailed image captions in text-to-image generation~\cite{dalle3}, we introduce the detailed description in our framework to enhance semantic similarity. However, the more words used, the higher the bitrates. To avoid unnecessary overhead and support flexible adjustment for users, we design a variable-length content prompt (ConP) with the interaction of a multimodal large language model~\cite{wang2023cogvlm}. ConP is divided into three levels based on the detail of the description as in Table~\ref{tab:illustration_of_ConP}, with each level having a length limit. Therefore, users can choose not to transmit text or to transmit text of different lengths at the encoder side according to the rate constraint, demonstrating the high flexibility of our framework. ConP will be losslessly compressed using Lempel-Ziv coding as implemented in the zlib library~\cite{zlib} similar to Lei \etal\cite{lei2023text} and PerCo~\cite{perco}. The average decoding time is less than 0.006 ms.

\myparagraph{Compression prompt.} To realize adaptive transformation for all basic codecs within one model, we design a compression prompt (ComP) that retains the information of the chosen basic codec and ConP. As shown in Fig~\ref{fig:overview}, ComP includes the following information: codec category, codec name, optimization metric, quality indicator, and ConP level. More concretely, the codec category is described as either traditional or neural. The optimization metric includes PSNR, MS-SSIM and GAN, with all traditional codecs defaulting to PSNR as their optimization metric. The quality indicator represents the configuration parameters related to the quality of codecs, such as the QP, Quantization Parameter in VTM~\cite{VVC}.

\subsection{Universal Perceptual Compensator}
\label{sec:Methods_Perceptual_Compensator}
After the transmission of textual prompts and compressed image $x_v$, the dual-modality information is fed into the proposed universal perceptual compensator to conduct conditional generation, which is based on the powerful text-to-image generation prior of Stable Diffusion~\cite{rombach2022high} (SD). Specifically, as illustrated in Fig.~\ref{fig:overview}, the semantic information ($C_{ConP}$) is fed into the perceptual compensator through cross-attention as in SD.
Moreover, motivated by the great success of adapters in controllable image generation~\cite{controlnet,t2iadapter}, we also input visual information into SD through the proposed universal visual adapter.
To condition it both on time step $t$ and the textual embedding of compression prompt extracted by a pre-trained CLIP text encoder ($C_{ComP}$), motivated by the global semantic embedding in~\cite{sd3}, we apply average pooling and 2-layer MLP (with SiLU activations and 1024 channels for each of the two dense layers on $C_{ComP}$ and add it with the transformed timestep $t$. Afterward, the resulting output is fed into each ResBlock of the adapter in the same manner as SD is conditioned on $t$. 
The final output of the universal visual adapter is then modulated into the Stable Diffusion model through the SPADE layer~\cite{wang2024exploiting,li2024ntire,ren2025moe}, which modulates the input feature using the provided scaling and shifting coefficients.

Besides, it is well-known that Stable Diffusion (SD) suffers from information loss due to the pretrained autoencoder, which severely affects the distortion performance of SD-based image codec~\cite{perco}. To address this problem, we finetune the VAE decoder $\mathcal{D}$ to improve the distortion after the diffusion UNet has been trained, helping our framework to cover a wider bitrate range. Except for the trainable decoder refiner, all other modules are frozen to save computational resources and preserve the generation prior in the pretrained autoencoder. We use the feature of compressed image $x_v$ in VAE encoder $\mathcal{E}$ to augment the original VAE decoder features through a skip connection. {Specifically, let $f_e$ represent the output feature of a layer in $\mathcal{E}$ and $f_{in}$ represent the symmetric layer's feature in $\mathcal{D}$. Within the Decoder Refiner module, $f_{in}$ and $f_e$ are first concatenated, and the resulting feature undergoes further processing through a series of residual blocks (ResBlocks) and an attention block to extract auxiliary feature, enhancing detail reconstruction before the residual connection. Finally, the refined features are merged with the original decoder feature $f_{in}$ through a residual connection.}
 
\begin{figure*}[t]
    \centering
    \begin{minipage}{0.245\textwidth}
        \centering
        \includegraphics[width=\linewidth]{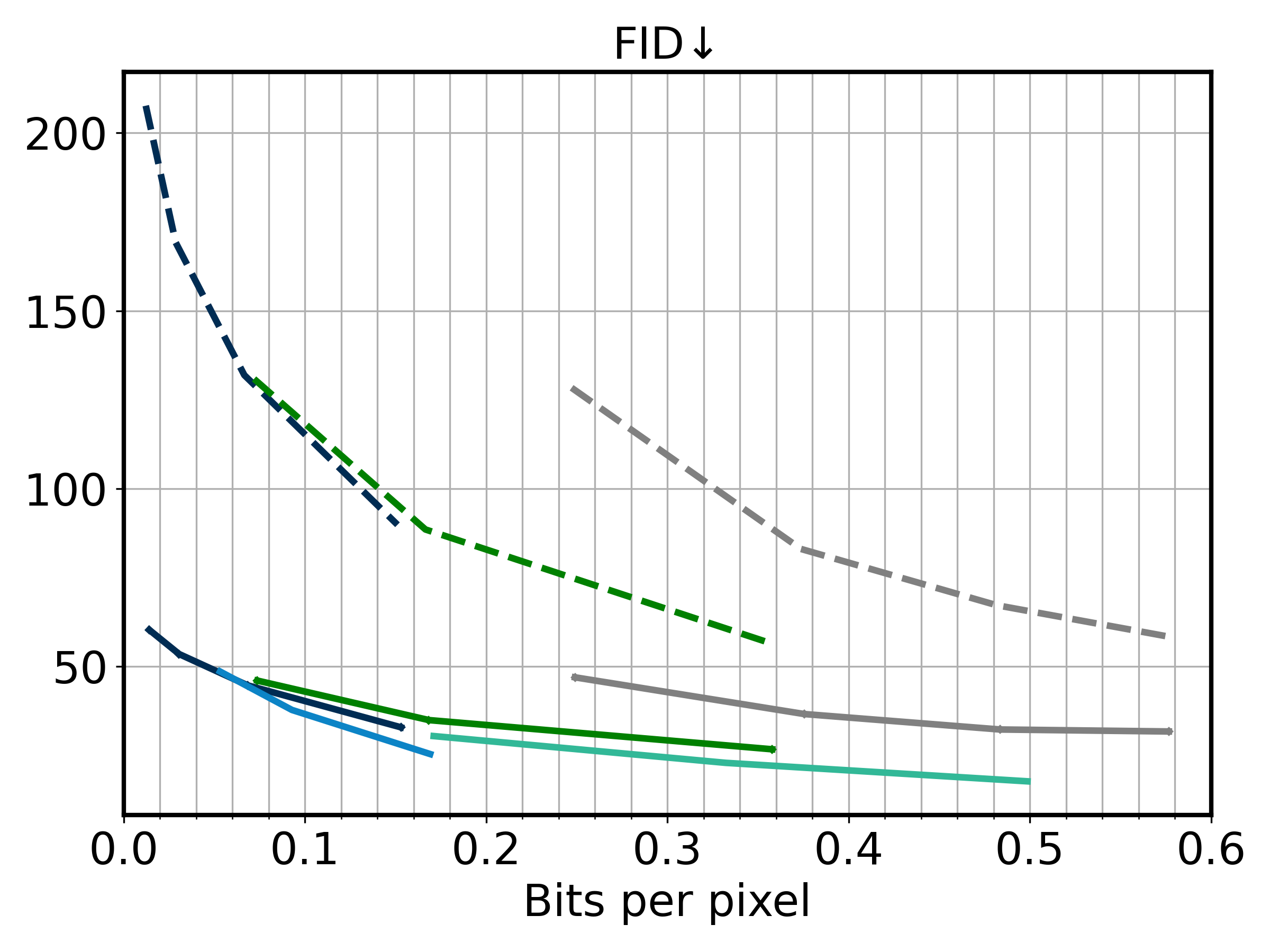}
    \end{minipage}
    \begin{minipage}{0.245\textwidth}
        \centering
        \includegraphics[width=\linewidth]{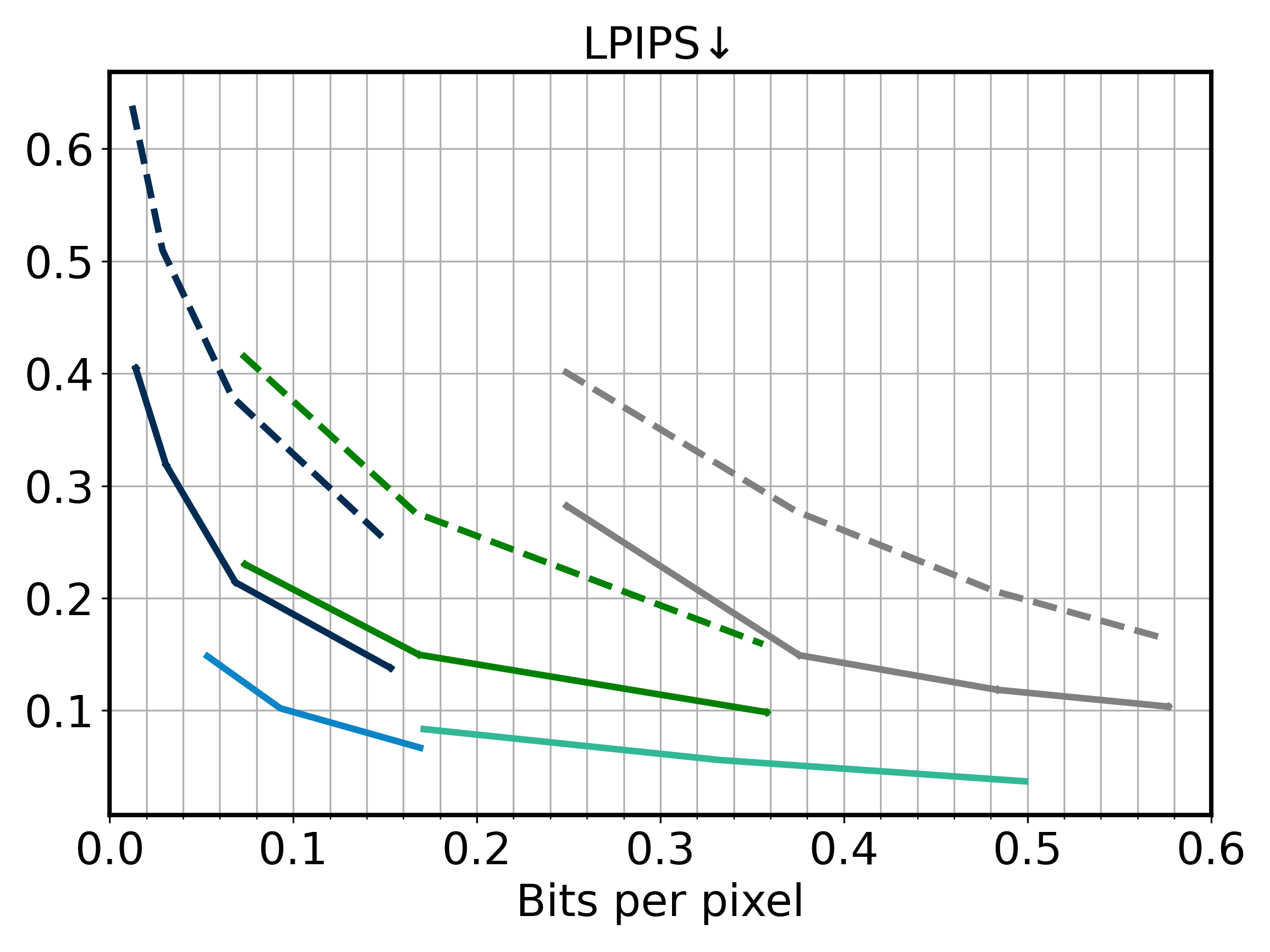}
    \end{minipage}
    \begin{minipage}{0.245\textwidth}
        \centering
        \includegraphics[width=\linewidth]{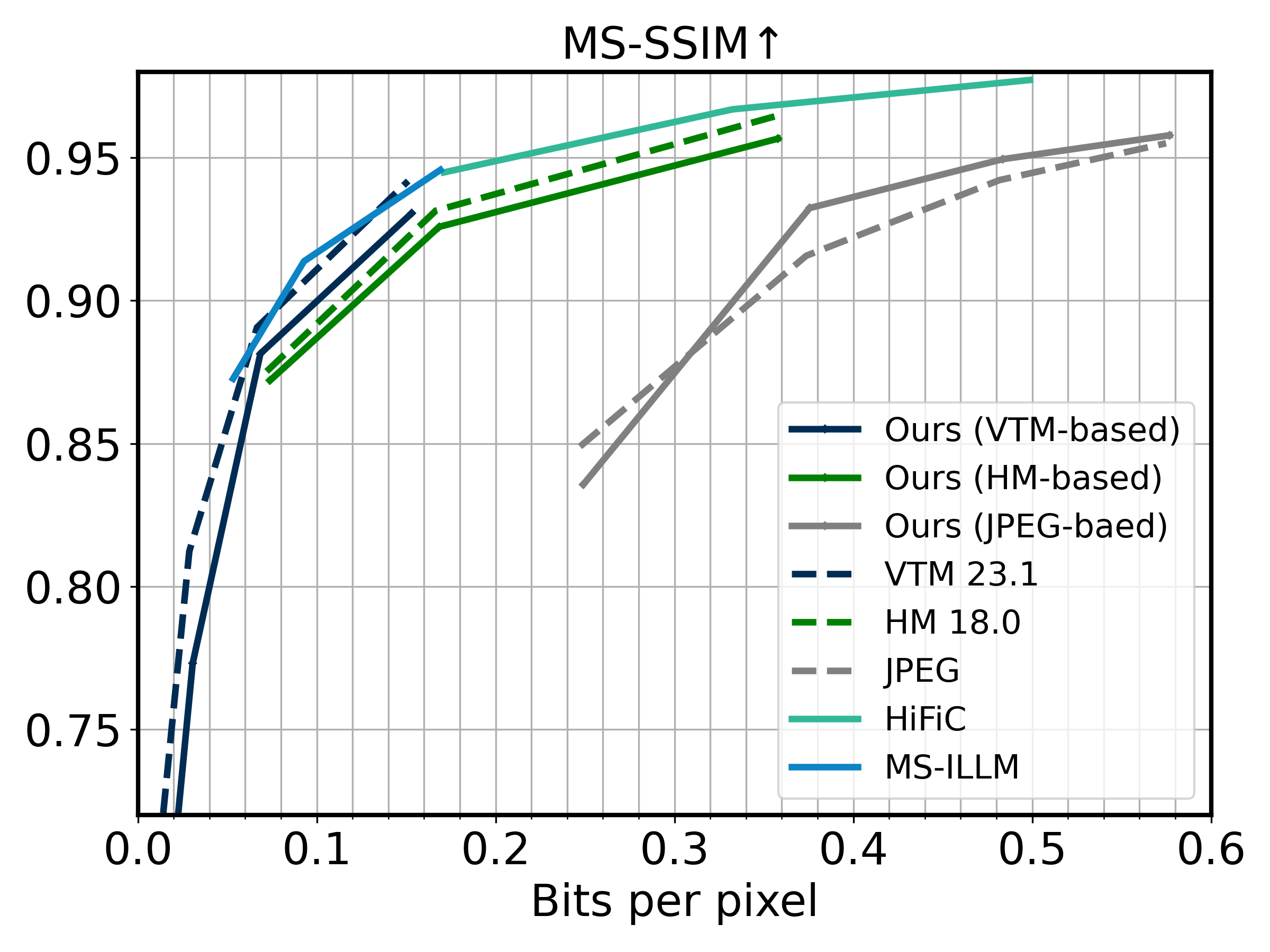}
    \end{minipage}
    \begin{minipage}{0.245\textwidth}
        \centering
        \includegraphics[width=\linewidth]{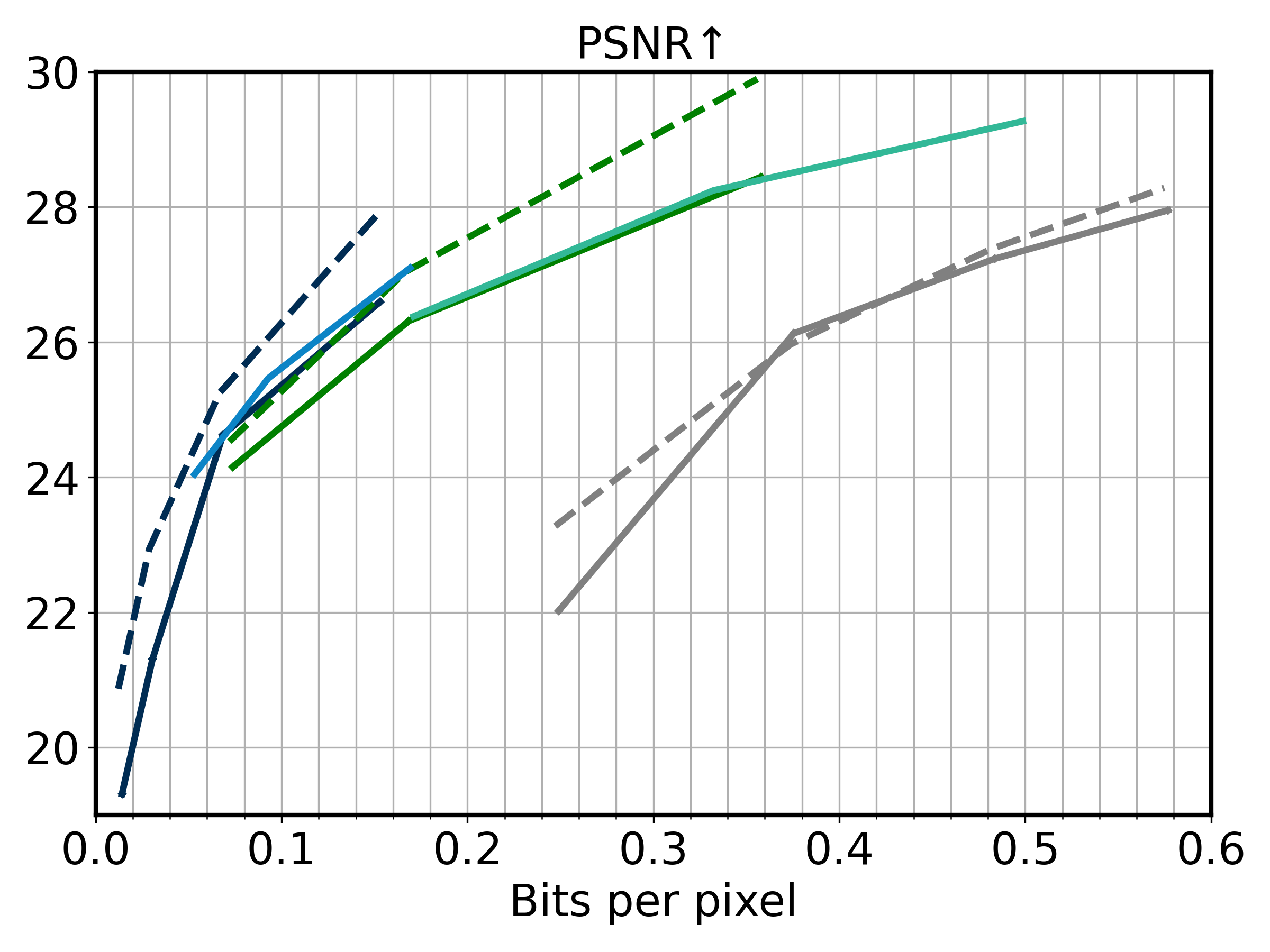}
    \end{minipage}
    \begin{minipage}{0.245\textwidth}
        \centering
        \includegraphics[width=\linewidth]{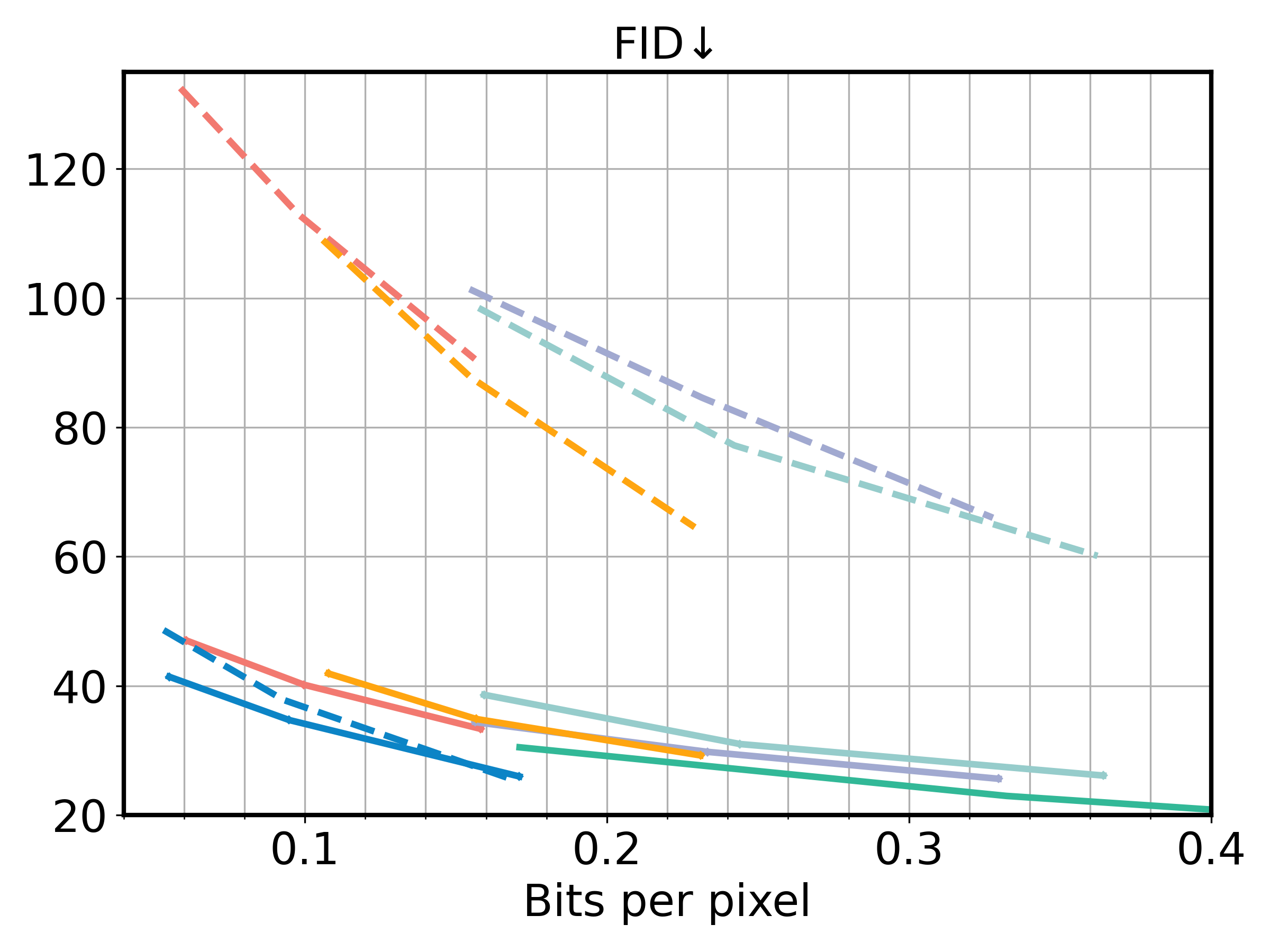}
    \end{minipage}
    \begin{minipage}{0.245\textwidth}
        \centering
        \includegraphics[width=\linewidth]{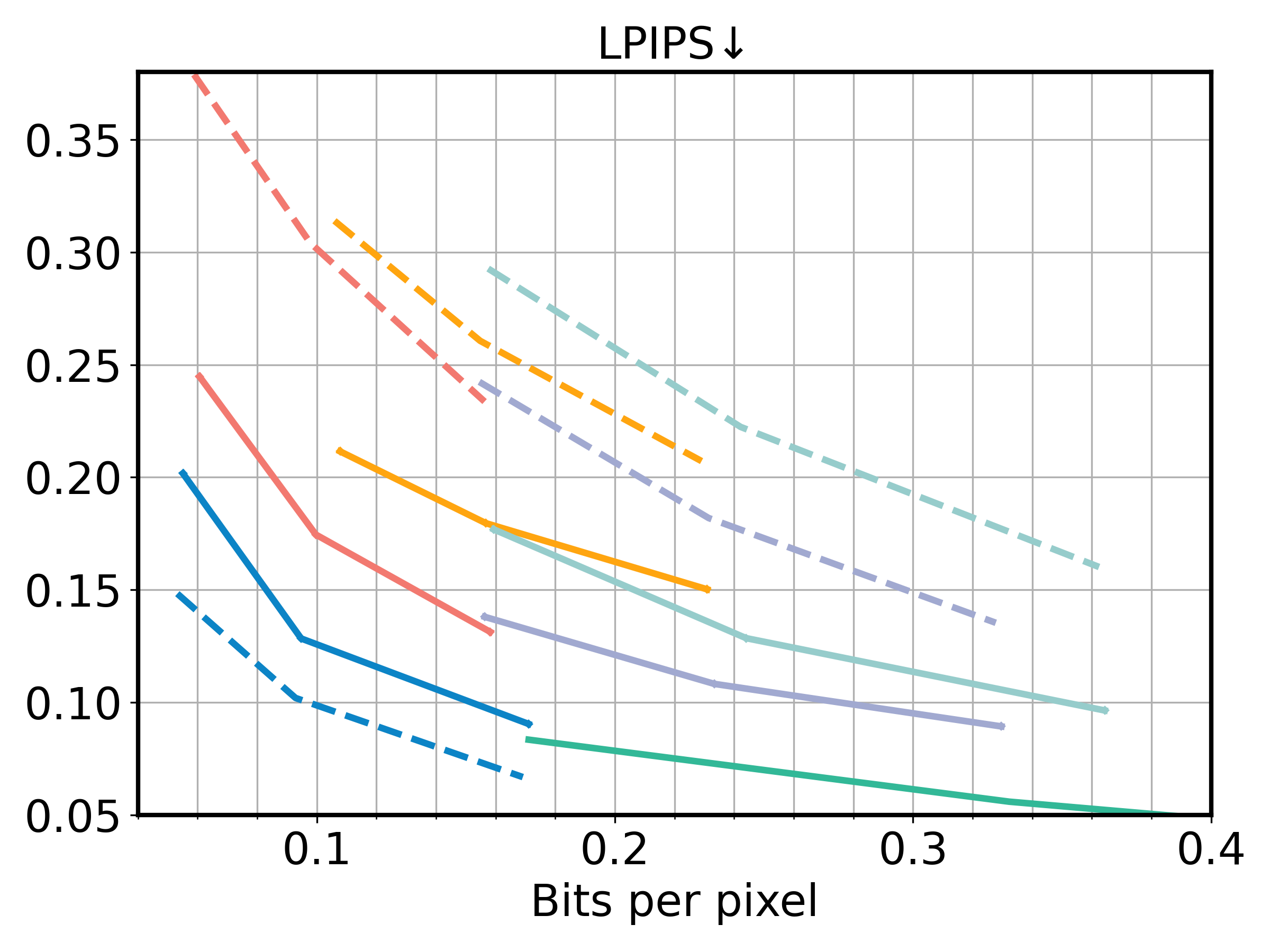}
    \end{minipage}
    \begin{minipage}{0.245\textwidth}
        \centering
        \includegraphics[width=\linewidth]{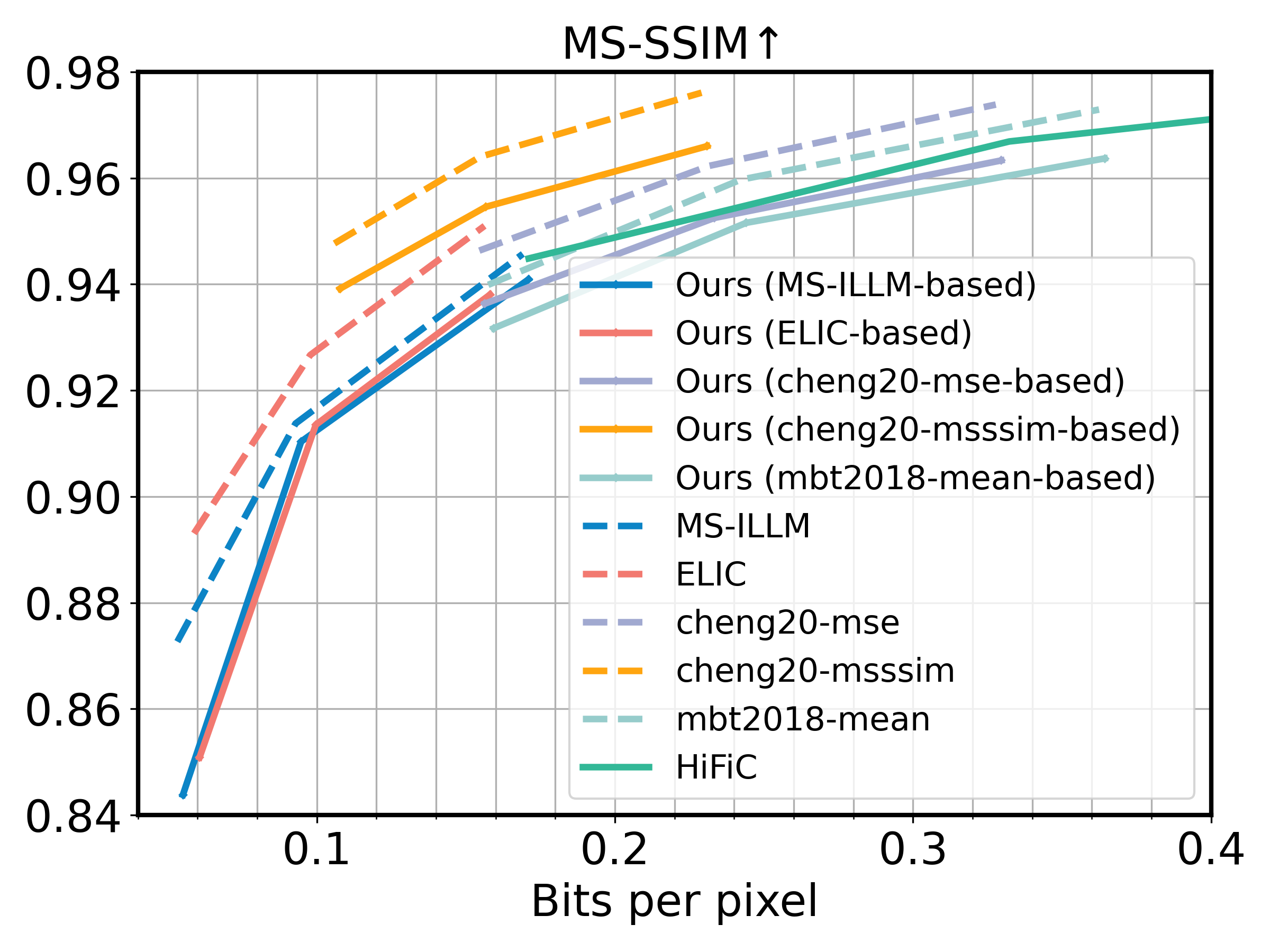}
    \end{minipage}
    \begin{minipage}{0.245\textwidth}
        \centering
        \includegraphics[width=\linewidth]{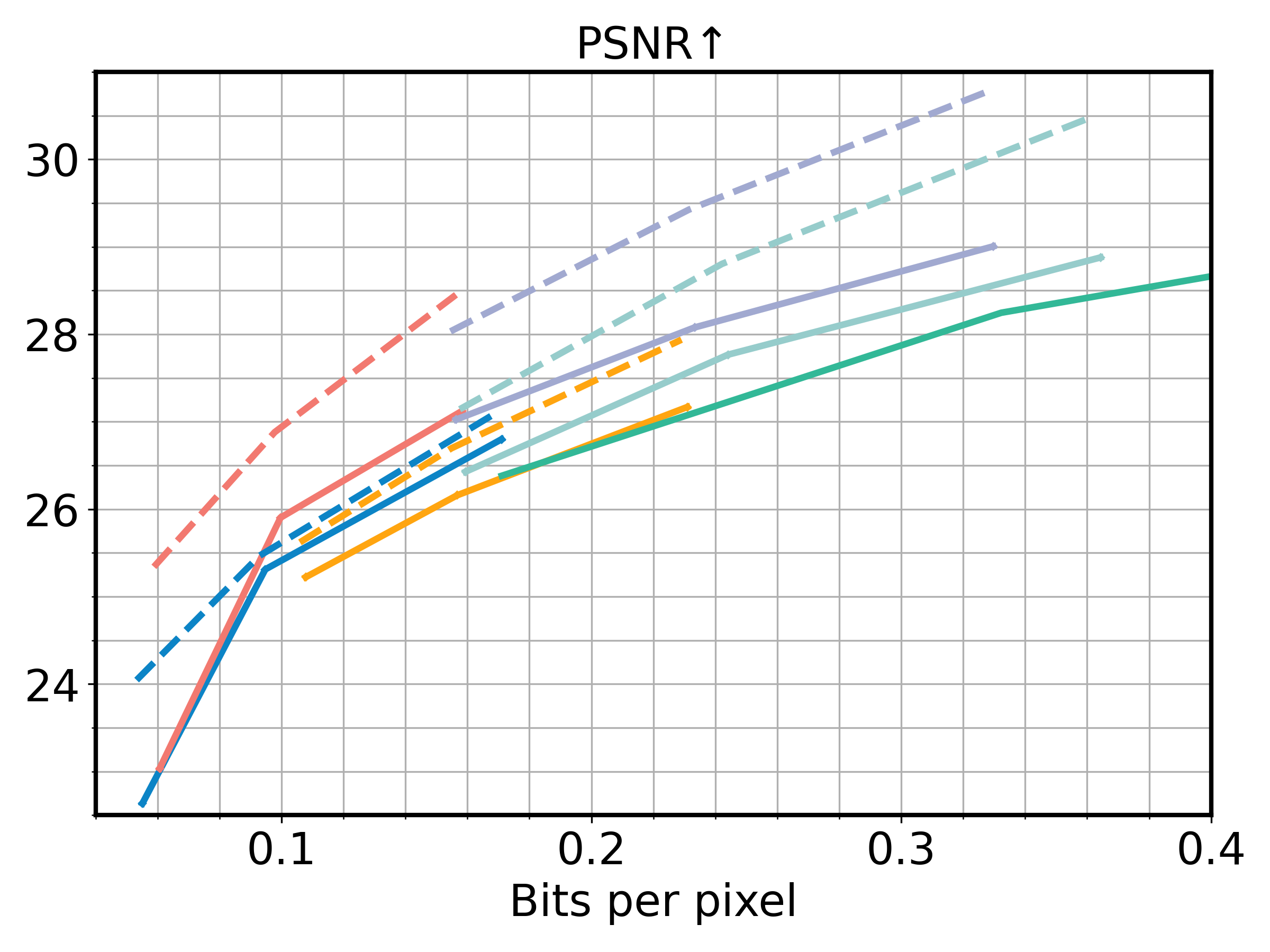}
    \end{minipage}
    \vspace{-2mm}
    \caption{Overall performance comparison between our method and state-of-the-art codecs on DIV2K.}
    \label{fig:main_results}
\end{figure*}
\subsection{Overall Framework}
\label{sec: Methods_OverallFramework}
The entire training process is divided into two stages. In the first stage, we only fine-tune the diffusion UNet with a few trainable parameters and the total trainable parameters constitute 12\% of all weights of diffusion UNet. Following Stable Diffusion, we also use classifier-free guidance for conditional generation and extend it to multi-conditional classifier-free guidance like ~\cite{avrahami2023spatext,brooks2023instructpix2pix} to adapt to two independent textual conditions. During training, we randomly set the probability of $C_{ComP} = \emptyset_{ComP}$ and $C_{ConP} = \emptyset_{ConP}$ to 10\% each. Therefore, the probability of not having any text conditional is 1\%. 
During sampling, for simplicity, we only calculate the direction of both conditions and extrapolate along this single direction. 
Formally, the training loss and sampling process follows Equation (\ref{eq: SD_train}) and Equation (\ref{eq: Classifier_sample}), respectively, where \( C = \{C_{ComP}, C_{ConP}\} \).
Both textual conditions can be selectively transmitted flexibly for various practical applications.
In the second stage, the training loss involves a combination of reconstruction (L1) loss, LPIPS~\cite{zhang2018unreasonable}, and adversarial loss~\cite{goodfellow2014generative}, similar to VQGAN~\cite{vqgan}.
\section{Experiments}
\label{sec:experiments}

\subsection{Experiment Setup}
\label{sec: Experiments_Experiment_Setup}
\myparagraph{Implementation details.} 
Our codec is built upon Stable Diffusion 2.1-base~\cite{rombach2022high}. For the highly descriptive content prompt generation, we use the state-of-the-art vision-language model, CogVLM~\cite{wang2023cogvlm}. The visual codec repository composes eight representative codecs in training: 1) traditional codecs: JPEG~\cite{wallace1992jpeg}, HM-intra 18.0~\cite{HEVC}, VTM-intra 23.1~\cite{VVC}. 2) neural image codecs: mbt2018-mean~\cite{minnen2018joint}, cheng20-anchor-mse~\cite{cheng2020learned}, cheng20-anchor-msssim, ELIC~\cite{he2022elic}, MS-ILLM~\cite{msillm}, where mbt2018-mean is Mean\&Scale version in~\cite{minnen2018joint} and cheng20-anchor is the varient model from~\cite{cheng2020learned} which has no self-attention. cheng20-anchor-mse means MSE-optimized cheng20-anchor and cheng20-anchor-msssim is MS-SSIM-optimized one. More details are included in the supplementary.

We train our model on Openimages~\cite{kuznetsova2020open} and resize them to $320\times320$ patches. Since we choose 3-4 bitrate points for each codec, this results in a 6.7M training set. For finetuning the VAE decoder, we generate 300K latent images with the weights from the first stage. We use Adam~\cite{kingma2014adam} optimizer and the learning rate is set to $5e-5$ all the time. 
The first stage involves 250K iterations, while the second stage involves 50K iterations.  The batch size is set to 32, and the training is conducted on four NVIDIA RTX 3090 GPUs. For inference, we use DDIM sampler~\cite{ddim} for 50 steps.

\myparagraph{Test hyperparameters.}
For the classifier guidance setting when using text prompts, we set $w=5.0$ for JPEG (QF$=5$), HM-intra 18.0, VTM-intra 23.1 (QP$\in\{57,52,47\}$), elic, cheng20-anchor-mse. For other bitrates and codecs, we set $w=7.5$.

\myparagraph{Datasets.} For evaluation, we use DIV2K~\cite{div2k} validation set (100 images) and CLIC2020~\cite{CLIC2020} test split (428 images). We resize the images in both datasets with the shorter dimension equal to 768px and then center-crop the image to a $768\times768$ squared shape, similar to Yang \etal\cite{yang2024lossy}.

\myparagraph{Metrics.} 
To comprehensively quantify the image quality, we select four metrics. Specifically, we use FID\cite{fid} to estimate the divergence between distributions of compressed images and original images. Following HiFiC~\cite{mentzer2020high}, we report patched FID by segmenting images into non-overlapping 256 × 256 resolution patches for small test sets ($\leq$ 100 images). Distortion measures are PSNR, MS-SSIM and LPIPS~\cite{zhang2018unreasonable}. 

\myparagraph{Baselines.} We include all codecs specified in the above and add HiFiC~\cite{mentzer2020high}, which is a state-of-the-art perceptual codec. For details on the replication of baselines please see appendix. For all methods, we set the random seed as 42 to avoid randomness.

\subsection{Main Results}
\begin{figure*}[htb]
    \captionsetup[subfigure]{labelformat=empty,font=tiny, aboveskip=0pt, belowskip=0pt}
    \centering
    \begin{minipage}[b]{0.135\textwidth}
        \centering
        \subcaption*{Original}
        \includegraphics[width=\textwidth]{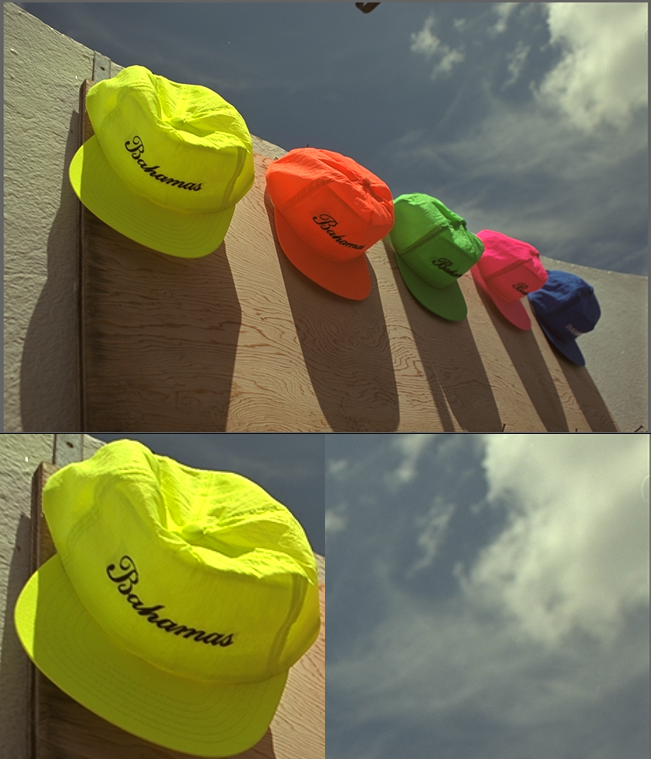}
        \subcaption{}
    \end{minipage}
    \begin{minipage}[b]{0.135\textwidth}
        \centering
        \subcaption*{VTM}
        \includegraphics[width=\textwidth]{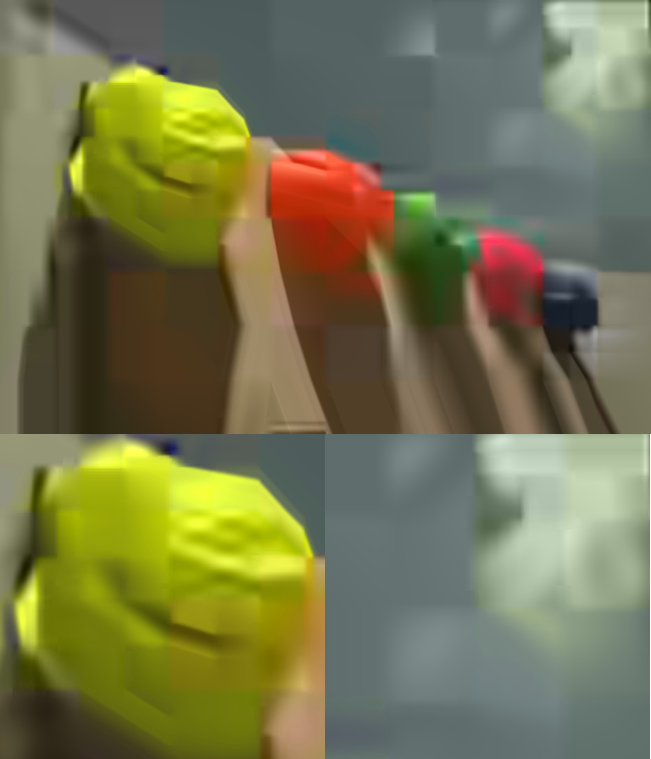}
        \subcaption{0.0073bpp}
    \end{minipage}
    \begin{minipage}[b]{0.135\textwidth}
        \centering
        \subcaption*{Ours w/o text}
        \includegraphics[width=\textwidth]{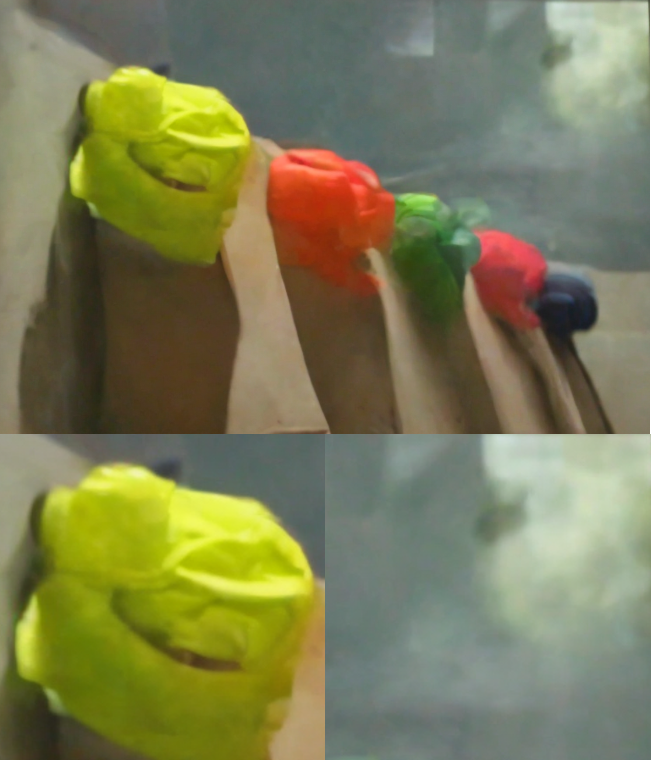}
        \subcaption{0.0073bpp}
    \end{minipage}
    \begin{minipage}[b]{0.135\textwidth}
        \centering
        \subcaption*{Ours w/o ConP}
        \includegraphics[width=\textwidth]{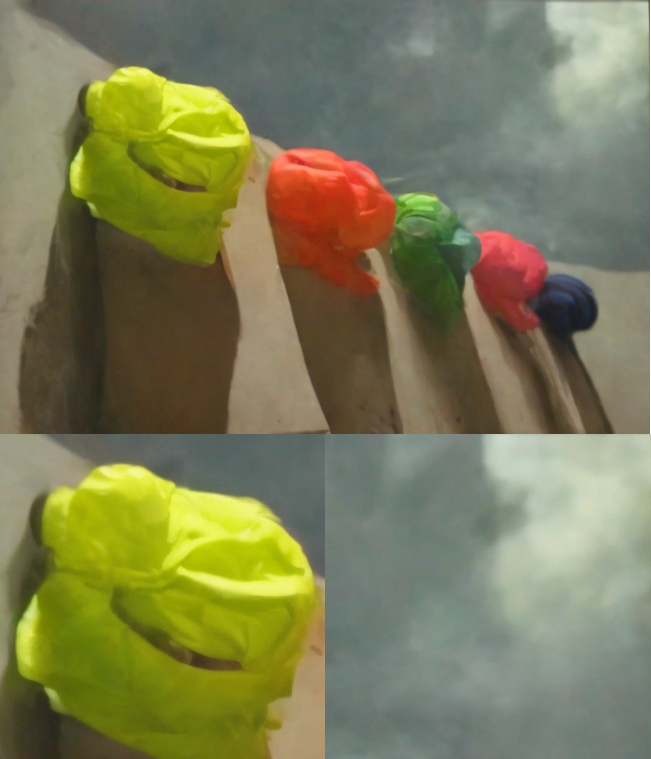}
        \subcaption{0.0090bpp}
    \end{minipage}
    \begin{minipage}[b]{0.135\textwidth}
        \centering
        \subcaption*{Ours w/ concise ConP}
        \includegraphics[width=\textwidth]{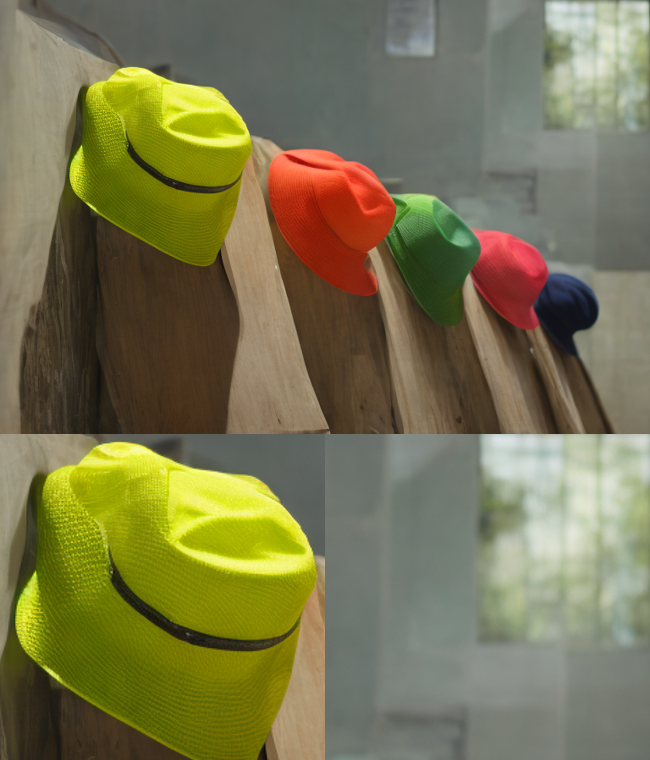}
        \subcaption{0.0106bpp}
    \end{minipage}
    \begin{minipage}[b]{0.135\textwidth}
        \centering
        \subcaption*{Ours w/ moderate ConP}
        \includegraphics[width=\textwidth]{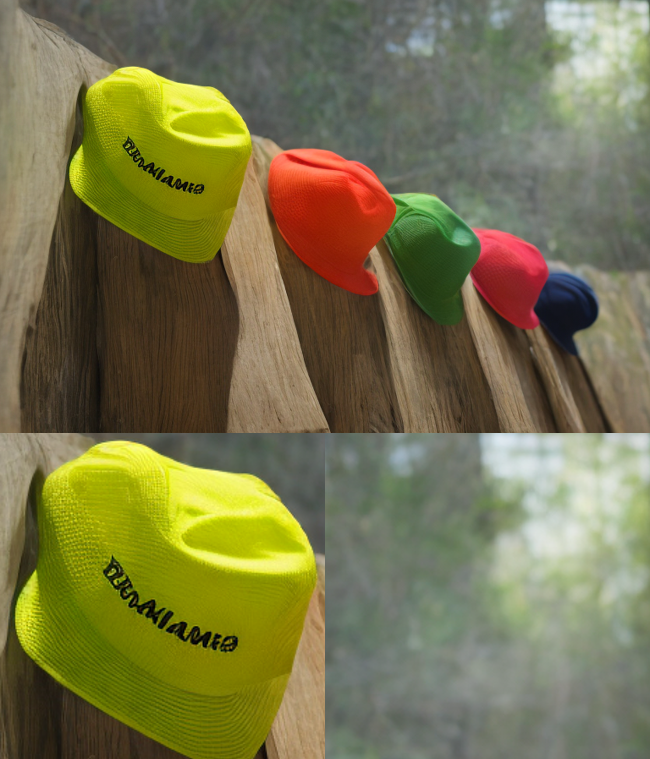}
        \subcaption{0.0119bpp}
    \end{minipage}
    \begin{minipage}[b]{0.135\textwidth}
        \centering
        \subcaption*{Ours w/ detailed ConP}
        \includegraphics[width=\textwidth]{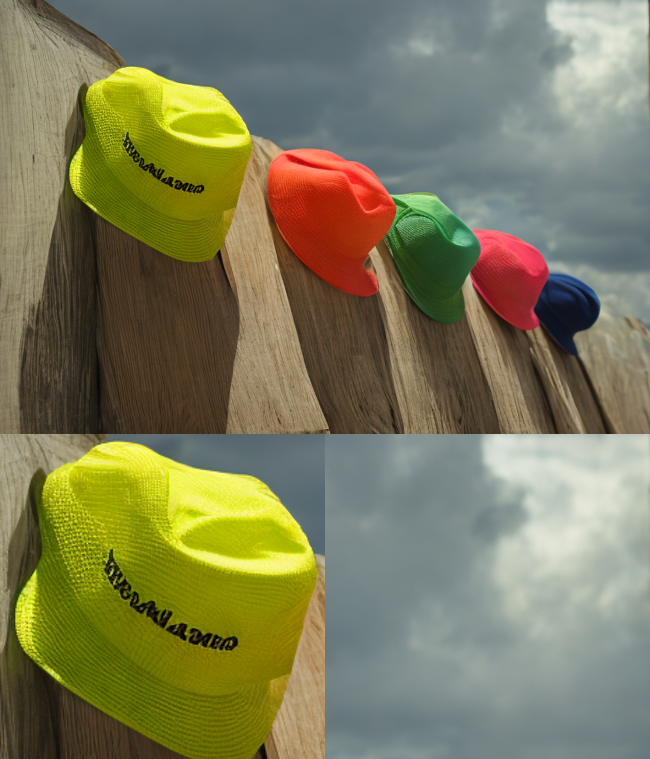}
        \subcaption{0.0131bpp}
    \end{minipage}
    \begin{minipage}[b]{0.135\textwidth}
        \centering
        \includegraphics[width=\textwidth]{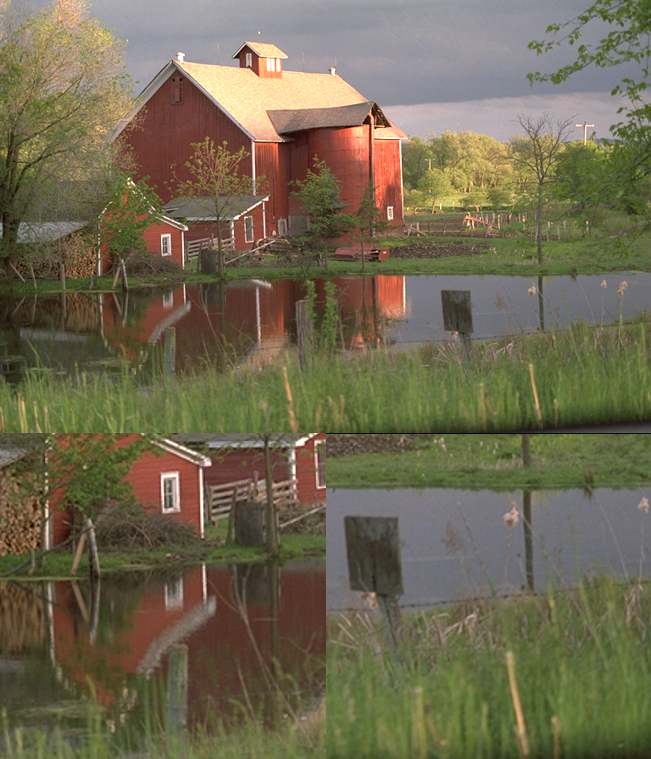}
        \subcaption{}
    \end{minipage}
    \begin{minipage}[b]{0.135\textwidth}
        \centering
        \includegraphics[width=\textwidth]{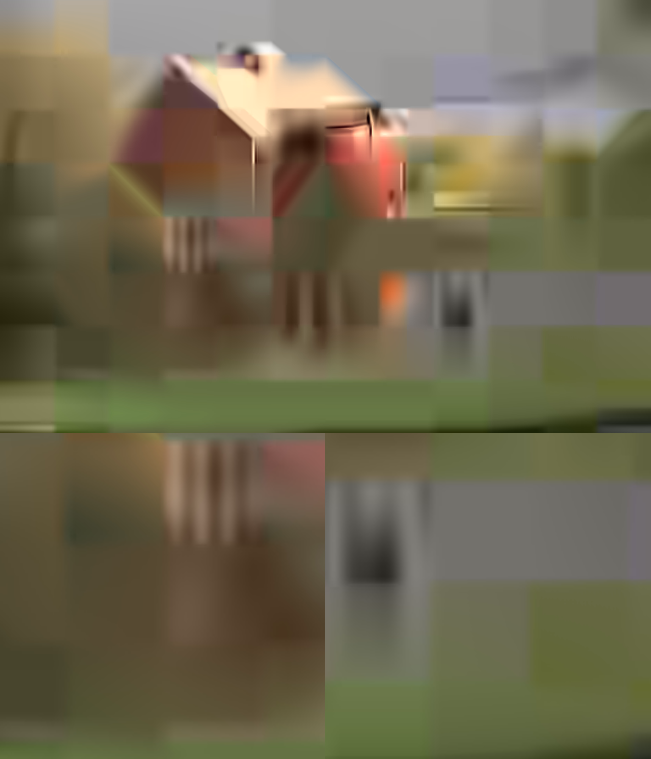}
        \subcaption{0.0064bpp}
    \end{minipage}
    \begin{minipage}[b]{0.135\textwidth}
        \centering        
        \includegraphics[width=\textwidth]{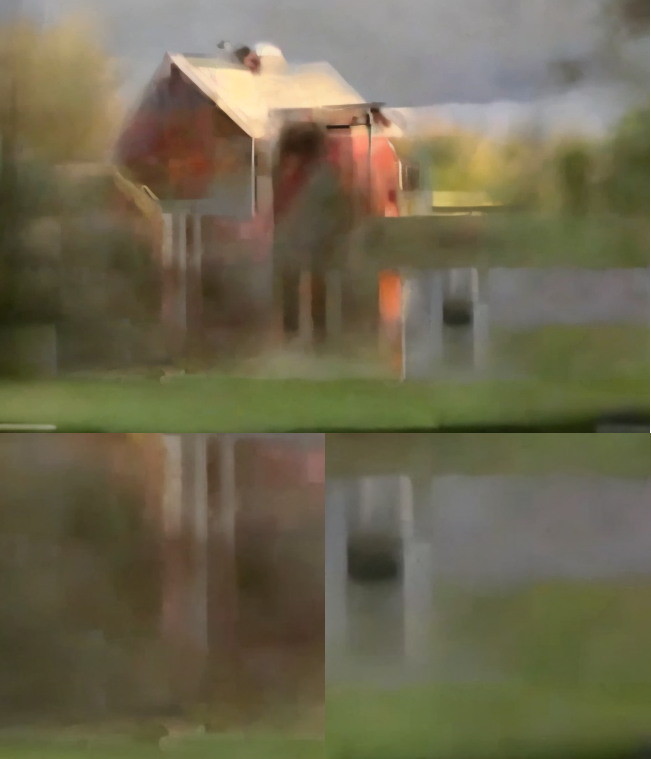}
        \subcaption{0.0064bpp}
    \end{minipage}
    \begin{minipage}[b]{0.135\textwidth}
        \centering
        \includegraphics[width=\textwidth]{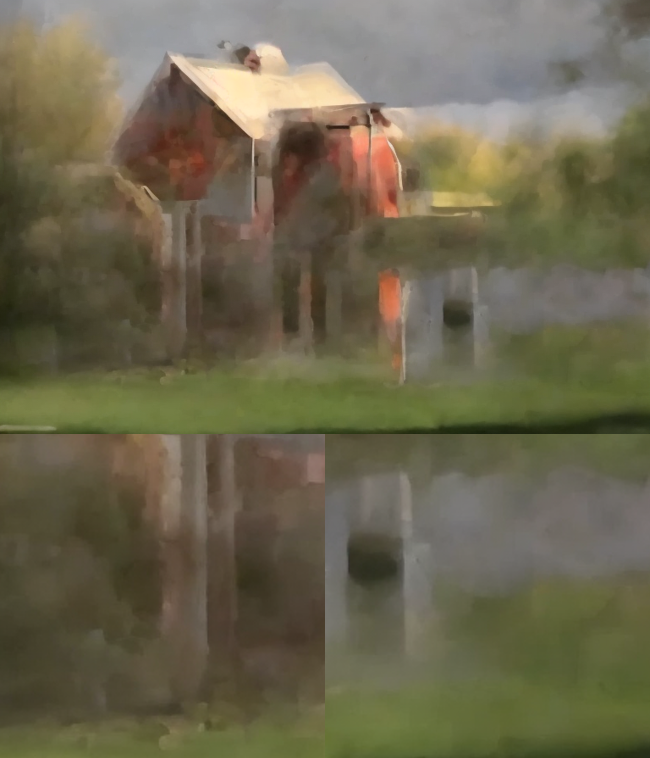}
        \subcaption{0.0081bpp}
    \end{minipage}
    \begin{minipage}[b]{0.135\textwidth}
        \centering
        \includegraphics[width=\textwidth]{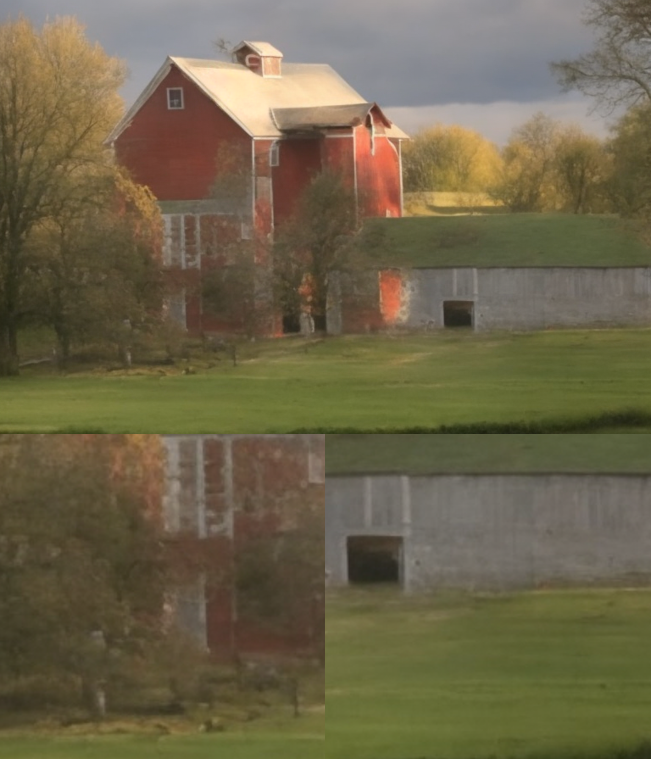}
        \subcaption{0.0094bpp}
    \end{minipage}
    \begin{minipage}[b]{0.135\textwidth}
        \centering
        \includegraphics[width=\textwidth]{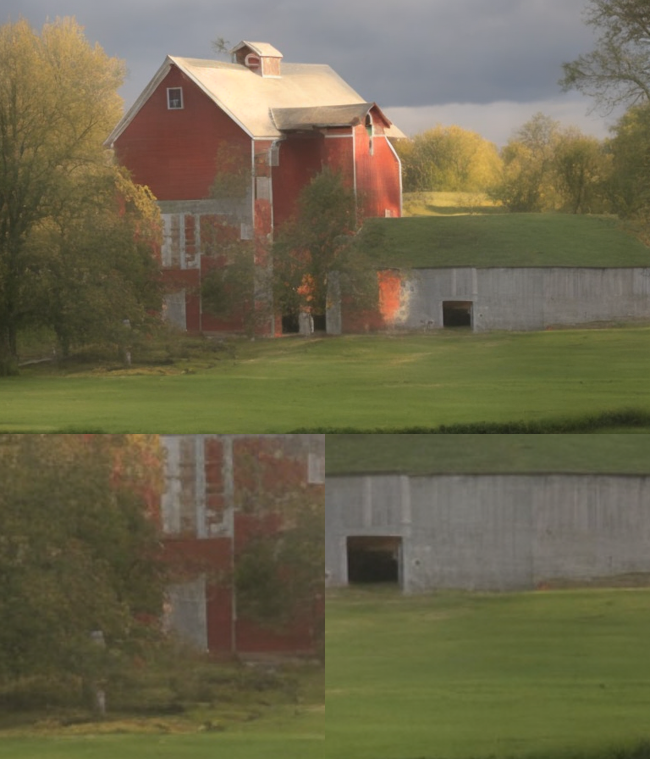}
        \subcaption{0.0102bpp}
    \end{minipage}
    \begin{minipage}[b]{0.135\textwidth}
        \centering
        \includegraphics[width=\textwidth]{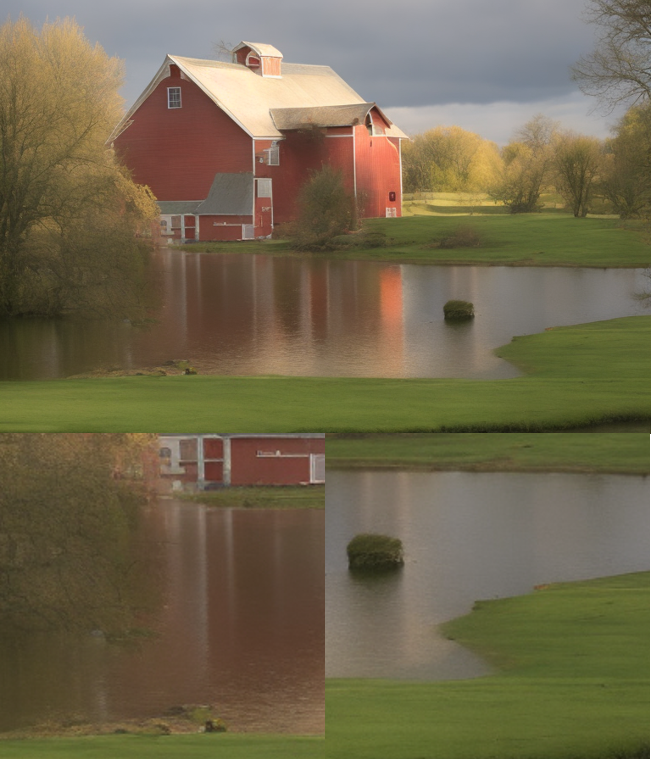}
        \subcaption{0.0131bpp}
    \end{minipage}
    \vspace{-2mm}
    \caption{Visual comparison of the reconstructions guided by different lengths of content prompt.}
    \label{fig:six_images}
\end{figure*}

\myparagraph{Quantitative results.} In Fig.~\ref{fig:main_results}, we compare our results to state-of-the-art (SOTA) codecs on DIV2K validation set. All of our models use detailed content prompts. 
We observe that our method achieves significant gains in terms of LPIPS and FID across all eight basic codecs, indicating the universality of our model.
Specifically, for traditional basic codecs (first row), our method achieves comparable FID values to the SOTA perceptual codec, MS-ILLM~\cite{msillm}, using VTM as our basic codec, effectively elevating it to a SOTA perceptual codec at ultra-low bitrates. For neural basic codecs, we achieve obvious FID improvements at lower bitrates ($\leq$0.15bpp) based on MS-ILLM, demonstrating our superiority in achieving high-fidelity results. 
While our UniMIC shows a drop in objective metrics like PSNR and MS-SSIM compared to the base codecs (except JPEG), this is due to the unavoidable trade-off between distortion and perceptual quality\cite{blau2019rethinking}. The results of CLIC2020 are included in the supplementary.

\myparagraph{Qualitative results.}
We compare our UniMIC with eight basic codecs in Fig.~\ref{fig:concept}. With a slight additional transmission overhead of textual prompts, our proposed approach achieves realistic and clear reconstructions. For example, leveraging the powerful generative prior of Stable Diffusion, we can turn a highly compressed VTM reconstruction, composed of some discontinuous mosaic blocks (first row, first column), into a semantically consistent (with original image) and artifacts-free high-definition image (first row, second column).
Besides, the result of MS-ILLM is filled with 'checkerboard' artifacts in this picture (zoom in for a clear comparison). In contrast, UniMIC (MS-ILLM-based) provides a much more clear and realistic reconstruction.

\myparagraph{Trade-off Between Distortion and Perception.}
To offer users greater flexibility, we provide a distortion-perception trade-off option in our method. 
As illustrated in paper~\cite{yan2021perceptual}, simple image interpolation between an MSE-optimized decoder and a perfect perceptual decoder is sufficient to achieve any point on the D-P curve. Inspired by this, we can interpolate between the codec and our perceptual result to realize the adjustable trade-off between distortion and perception. 
The equation is as follows:
\begin{equation}
\bar{x} = (1 - \beta) \times x_v +  \beta \times \hat{x}
\end{equation}
where $x_v$ is the output image of basic codec and $\hat{x}$ represents the output of our method. $\beta \in [0,1]$ is used to adjust the balance between distortion and perception. As $\beta$ approaches 1, the distortion of $\bar{x}$ increases, and the perception performance of $\bar{x}$ is enhanced. Conversely, as $\beta$ approaches 0, the distortion decreases, and the perception performance diminishes.

In Fig.~\ref{fig:six_images}, we present a visualization of images generated with varying $\beta$ values. As the $\beta$ value increases, the high-frequency information in the images progressively increases, and distortions such as blur and blocking are significantly alleviated. Therefore, users can flexibly adjust the degree of distortion and perception according to their preferences.

\begin{figure}[htb]
\centerline{\includegraphics[width=0.7\linewidth]{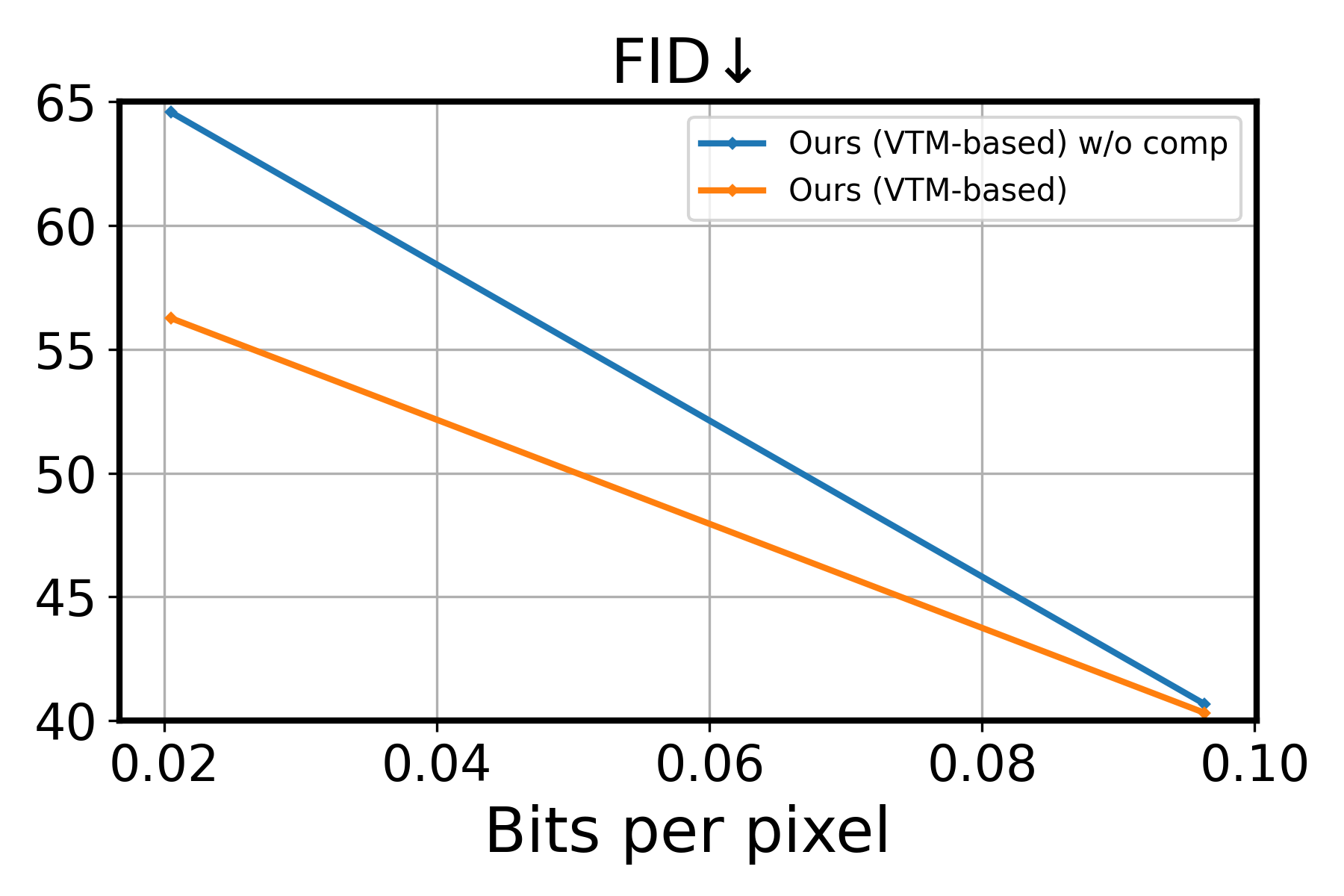}}
    \caption{Test our UniMIC on unseen bitrates in the training stage. }
\label{fig:unseen}
\end{figure}

\myparagraph{Extensible bitrate adaptation.}
Many image codecs support Variable-bit-rate, enabling bit adaptation within a single model (\egno, VTM, HM). This requires our model to be robust in adapting to unseen bitrates of basic visual codecs in the training stage. To verify this, we use VVC with QP values ranging from $\left[ 55,45 \right]$. To adapt to unseen bitrates, we perform interpolation between two seen prompts. For example, when adapting to VVC with QP 55, we generate a compression prompt by linearly interpolating between the prompts for QP 57 and QP 52. Fig.~\ref{fig:unseen} demonstrates that our model exhibits great adaptability to unseen bitrates during the training stage.

\begin{figure*}[t]
    \captionsetup[subfigure]{labelformat=empty,font=tiny, aboveskip=0pt, belowskip=0pt}
    \centering
    \begin{minipage}{0.15\textwidth}
        \centering
        \subcaption*{Original}
        \includegraphics[width=\textwidth]{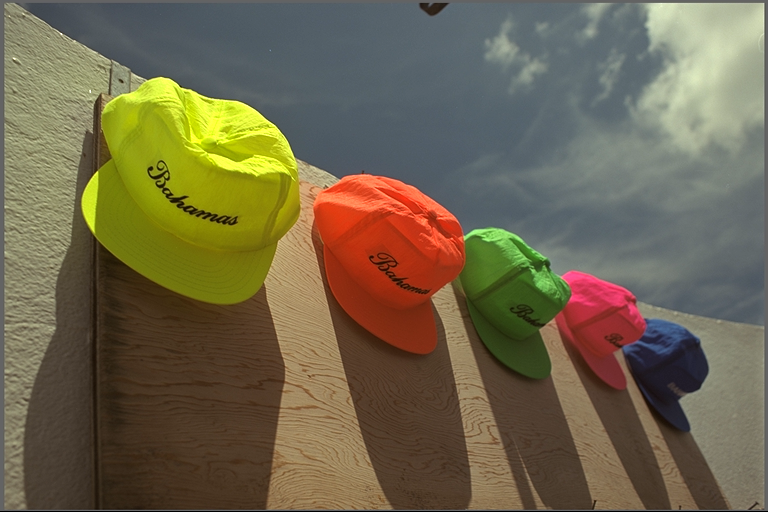}
    \end{minipage}
    \begin{minipage}{0.15\textwidth}
        \centering
        \subcaption*{VTM}
        \includegraphics[width=\textwidth]{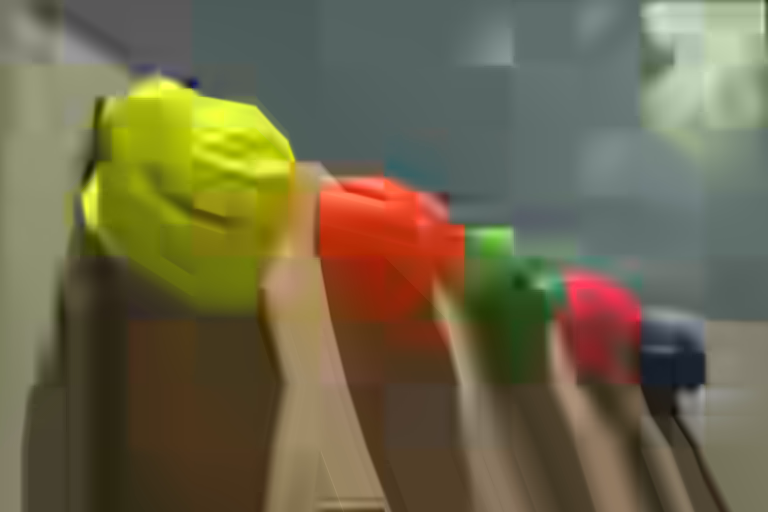}
    \end{minipage}
    \begin{minipage}{0.15\textwidth}
        \centering
        \subcaption*{$\beta = 0.2$}
        \includegraphics[width=\textwidth]{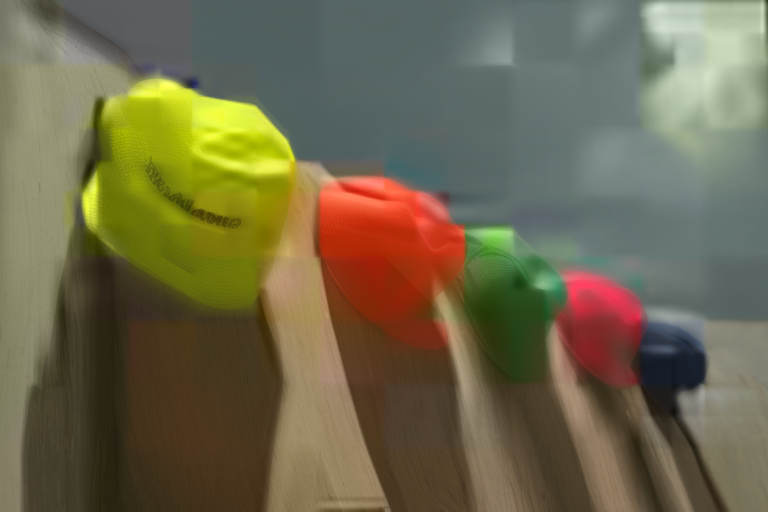}
    \end{minipage}
    \begin{minipage}{0.15\textwidth}
        \centering
        \subcaption*{$\beta = 0.5$}
        \includegraphics[width=\textwidth]{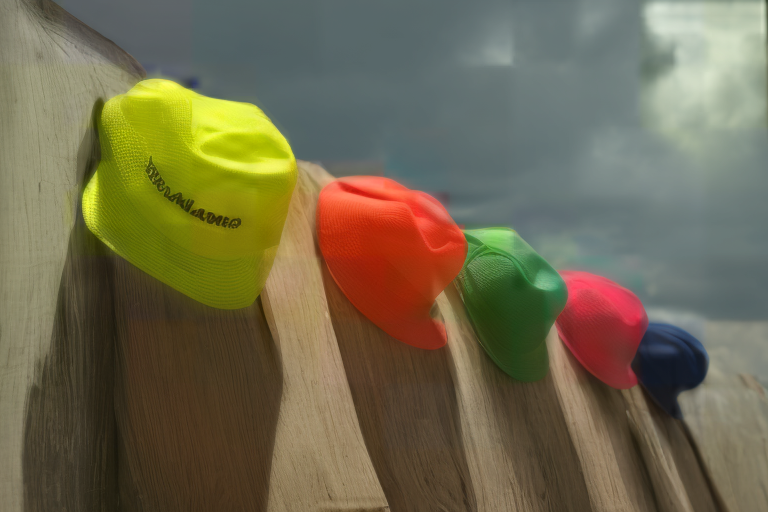}
    \end{minipage}
    \begin{minipage}{0.15\textwidth}
        \centering
        \subcaption*{$\beta = 0.8$}
        \includegraphics[width=\textwidth]{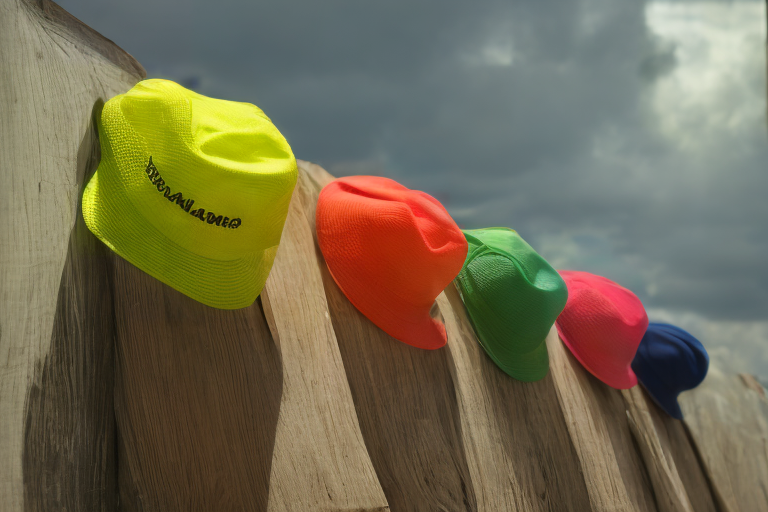}
    \end{minipage}
    \begin{minipage}{0.15\textwidth}
        \centering
        \subcaption*{Ours ($\beta = 1.0$)}
        \includegraphics[width=\textwidth]{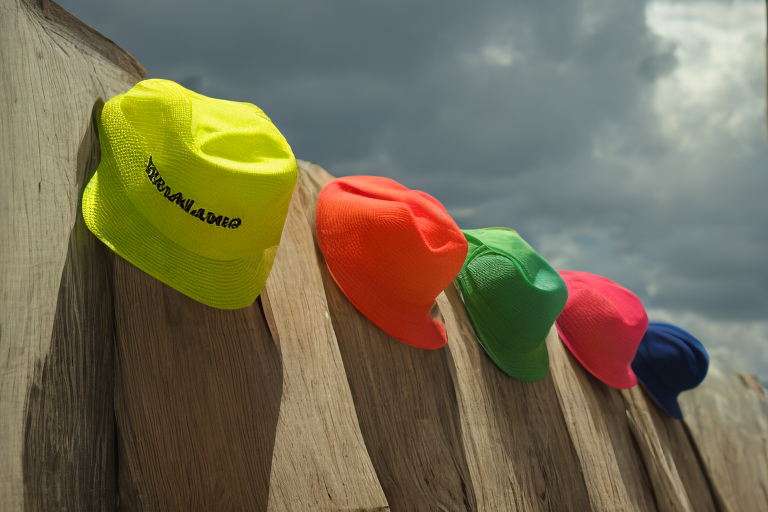}
    \end{minipage} \\
    \begin{minipage}{0.15\textwidth}
        \centering
        \includegraphics[width=\textwidth]{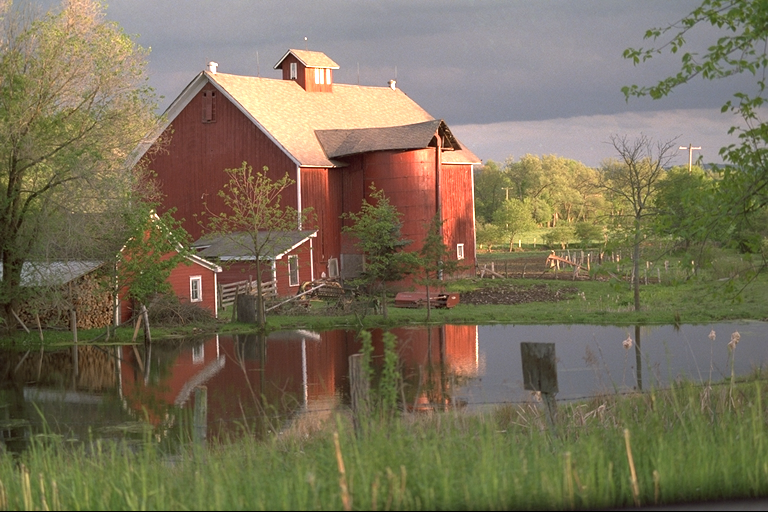}
    \end{minipage}
    \begin{minipage}{0.15\textwidth}
        \centering
        \includegraphics[width=\textwidth]{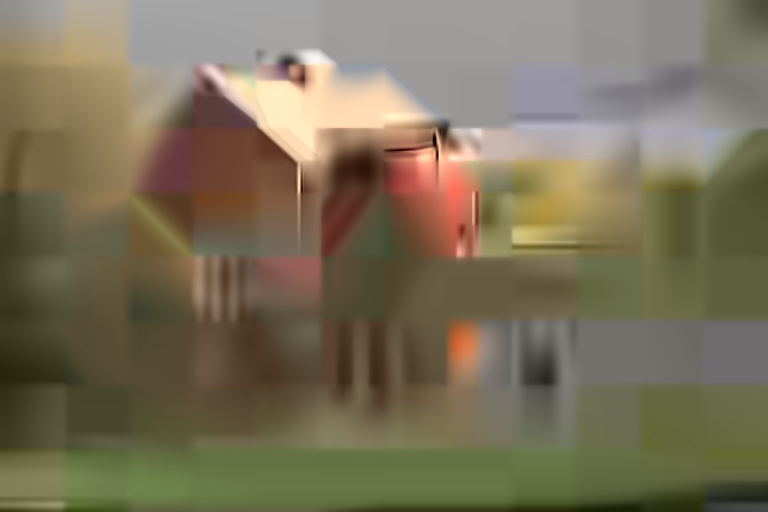}
    \end{minipage}
    \begin{minipage}{0.15\textwidth}
        \centering        
        \includegraphics[width=\textwidth]{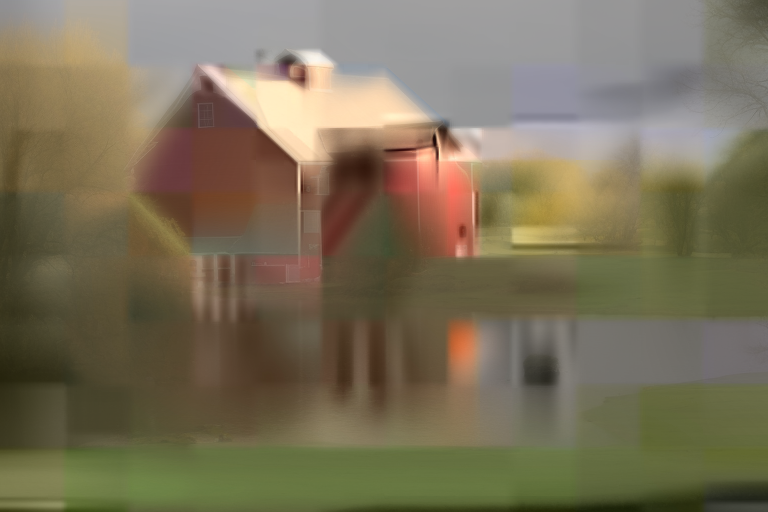}
    \end{minipage}
    \begin{minipage}{0.15\textwidth}
        \centering
        \includegraphics[width=\textwidth]{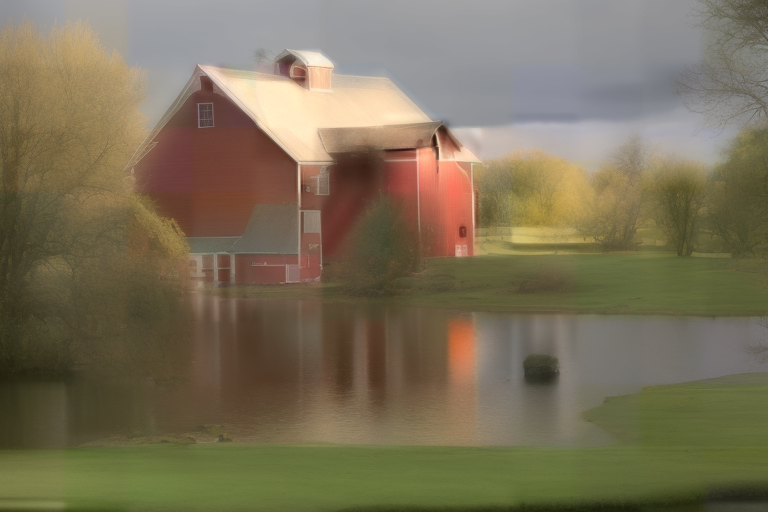}
    \end{minipage}
    \begin{minipage}{0.15\textwidth}
        \centering
        \includegraphics[width=\textwidth]{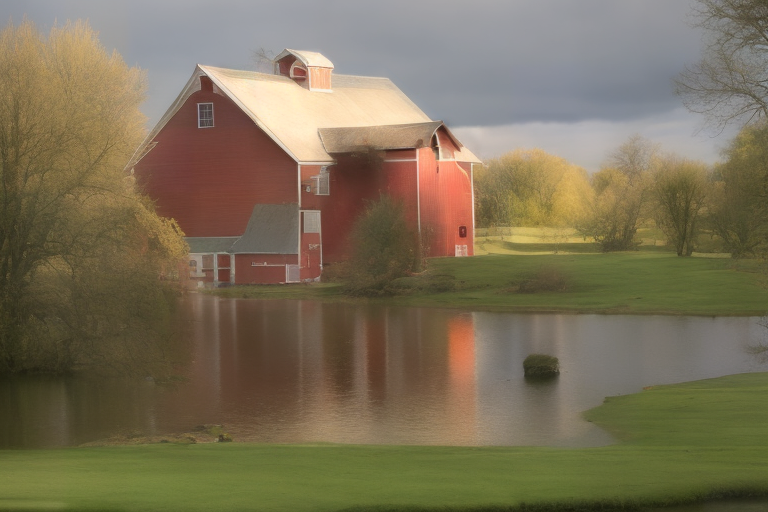}
    \end{minipage}
    \begin{minipage}{0.15\textwidth}
        \centering
        \includegraphics[width=\textwidth]{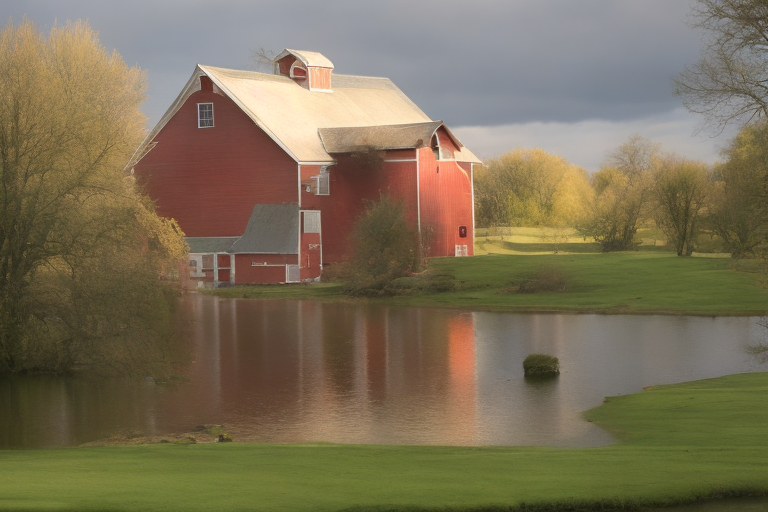}
    \end{minipage}
    \vspace{-2mm}
    \caption{Visualization results with different $\beta$ values.}
    \label{fig:six_images}
\end{figure*}

\myparagraph{Complexity}
The average decoding time for the 24 images in the Kodak dataset is 13.69 seconds. 
The total model size of our universal perceptual compensator is 1.45B parameters, with the pretrained Stable Diffusion model accounting for 949M of them.

\subsection{Ablation Study}
To investigate the effectiveness of our proposed multi-grained textual coding, we conduct experiments on the DIV2K dataset with the representative traditional SOTA codec VTM~\cite{VVC} and neural SOTA codec MS-ILLM~\cite{msillm} as our basic codecs.


\myparagraph{Effectiveness of textual coding.} 
Our ablation study results on textual prompts are shown in Tab. \ref{tab:2}. Compared to the baseline without any textual information, the introduction of Compression Prompt (ComP) effectively strengthens the adaptation of our model to different basic codecs. The use of the Content Prompt (ConP) results in more realistic reconstructions by providing semantic information.

\myparagraph{Effectiveness of variable-length Content Prompt.}
\begin{table}[t]
    \centering
    \resizebox{0.48\textwidth}{!}{
    \begin{tabular}{ccccc}
    \toprule
    Base Codec &  \begin{tabular}[c]{@{}c@{}}Ours\end{tabular} & \begin{tabular}[c]{@{}c@{}}Ours w/o ComP\end{tabular} & \begin{tabular}[c]{@{}c@{}}Ours w/o ConP\end{tabular} & \begin{tabular}[c]{@{}c@{}}Ours w/o text \end{tabular} \\
    \midrule
    VTM & -96.91 & -94.61 & -85.06 & -83.60 \\
    MS-ILLM & -3.71 & -2.95 & -1.14 & -0.96 \\
    \bottomrule
    \end{tabular}}
    \vspace{-2mm}
    \caption{Ablation of textual prompts. The evaluation metric is BD-FID(\%) (lower is better). The values are calculated between the results of each base codec and our UniMIC.}
    \label{tab:2}
\end{table}
\begin{figure}[t]
    \vspace{-4mm}
    \centering
    \begin{minipage}{0.22\textwidth}
        \centering
        \includegraphics[width=\linewidth]{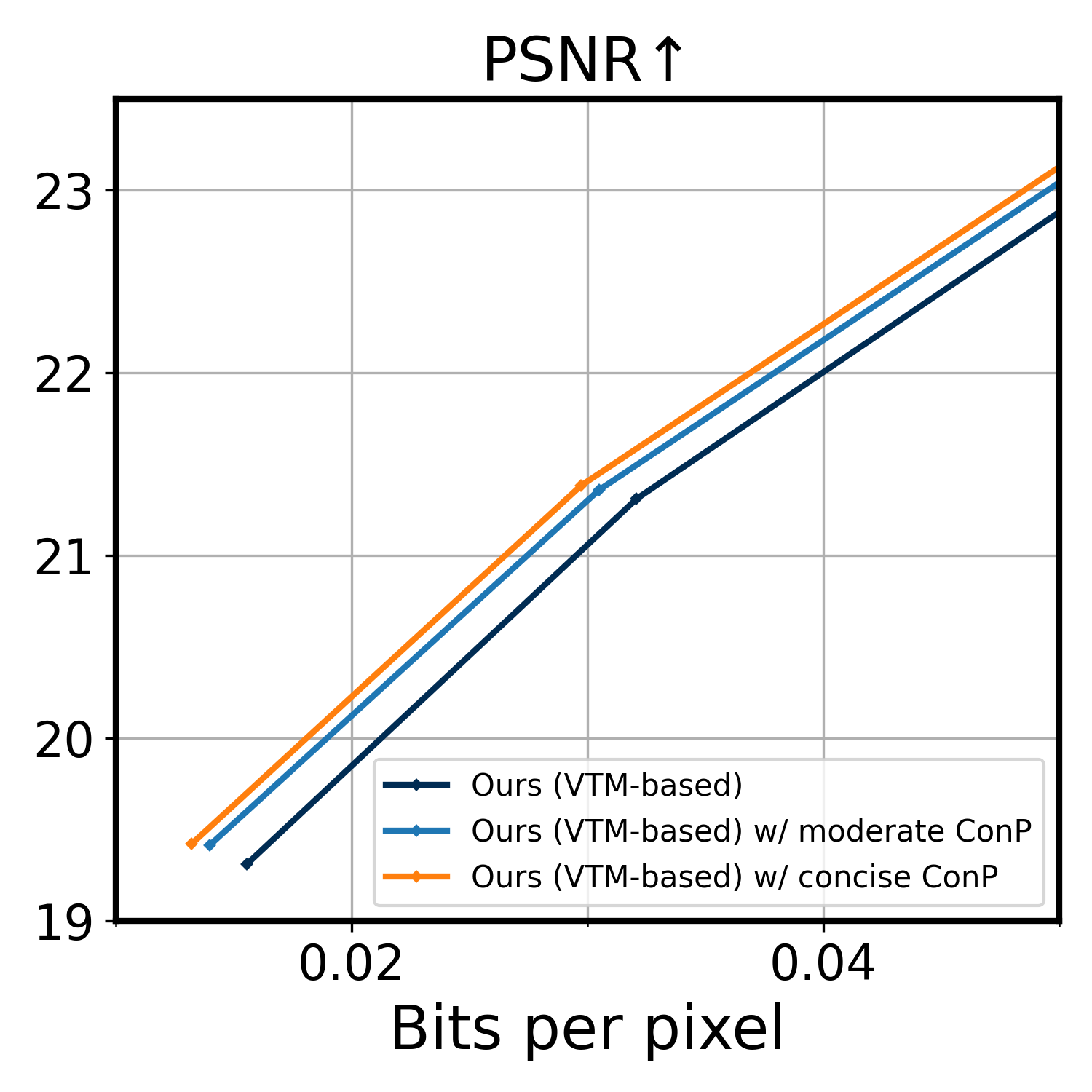}
    \end{minipage}
    \begin{minipage}{0.22\textwidth}
        \centering
        \includegraphics[width=\linewidth]{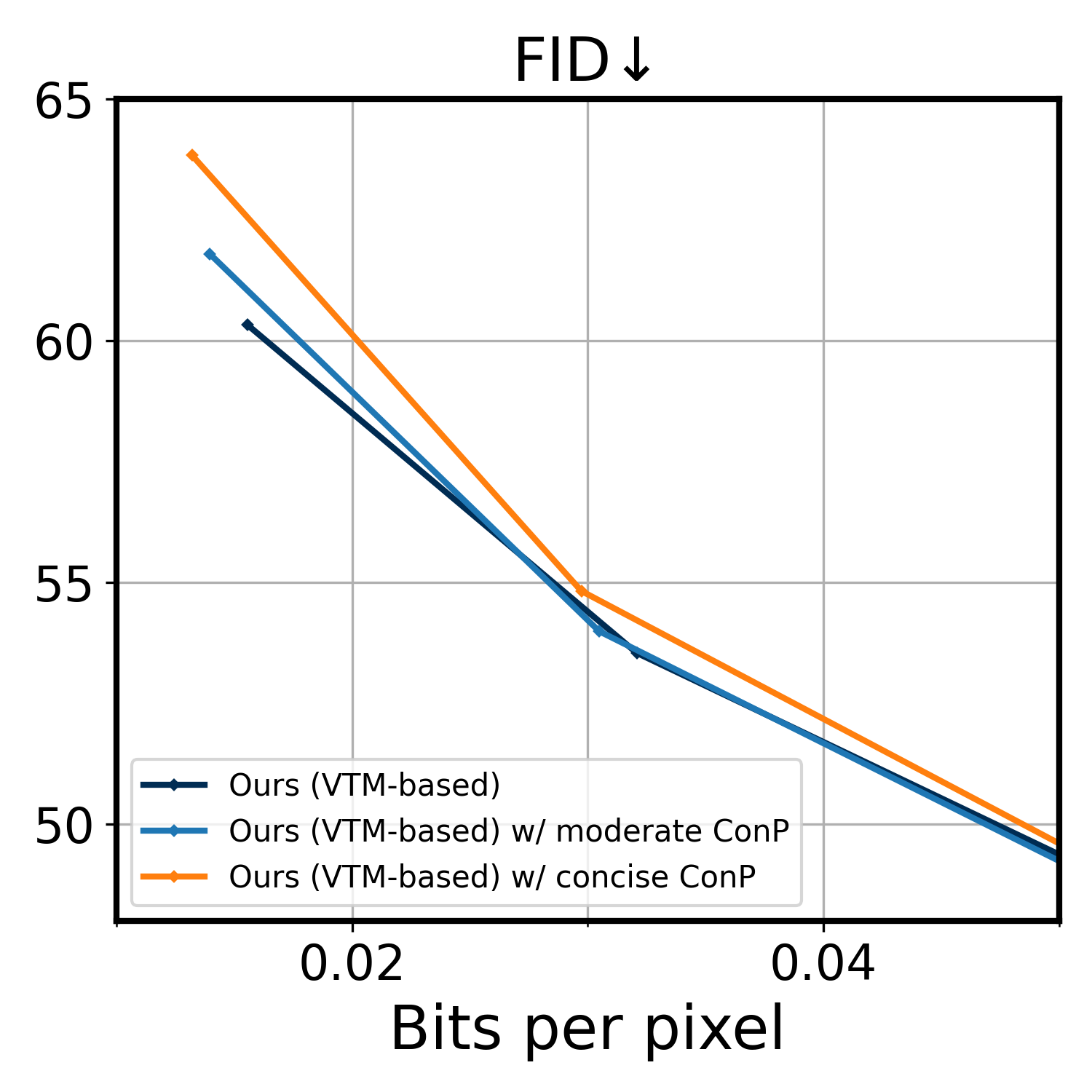}
    \end{minipage}
    \vspace{-2mm}
    \caption{Abalation of variable-length content prompt.}
    \label{fig:ablation_sp}
\end{figure}
The effect of variable-length ConP is illustrated in Fig~\ref{fig:ablation_sp}. As the length of ConP increases, there is a general trend of greater distortion but better perceptual quality. However, the differences between varying lengths gradually diminish as the amount of visual representation information increases. Therefore, we recommend using detailed descriptions at lower bitrates to achieve more realistic results. Fig~\ref{fig:six_images} presents qualitative results of varying text prompts. As the amount of textual information increases, the images transition from being blurry and blocky to clear and perceptually pleasing.

\myparagraph{Effectiveness of Decoder Refiner.}
\begin{table}[t]
    \vspace{-2mm}
    \centering
    \tiny
    \resizebox{0.38\textwidth}{!}{
    \begin{tabular}{ccc}
    \toprule
    Base Codec &  \begin{tabular}[c]{@{}c@{}}Ours\end{tabular} & \begin{tabular}[c]{@{}c@{}}Ours w/o Decoder Refiner\end{tabular}\\
    \midrule
    VTM & -96.91 / -1.40 & -96.76 / -2.36 \\
    MS-ILLM & -3.71 / -0.62 & -2.88 / -1.66 \\
    \bottomrule
    \end{tabular}}
    \vspace{-2mm}
    \caption{Ablation of the usage of decoder refiner. The evaluation metrics are presented as BD-FID(\%) (lower is better) / BD-PSNR(\%) (higher is better). The values are calculated between the results of each base codec and our UniMIC.}
    \label{tab:haha}
\end{table}
In Fig~\ref{tab:haha}, it is evident that our method significantly enhances fidelity in terms of objective
metrics. This improvement is primarily due to the information loss in the pretrained autoencoder of Stable Diffusion. By incorporating the decoder refiner, the supported bitrate range of our model is further extended.

\begin{figure}[htb]
    \centering
    \begin{minipage}{0.18\textwidth}
        \centering
        \includegraphics[width=\linewidth]{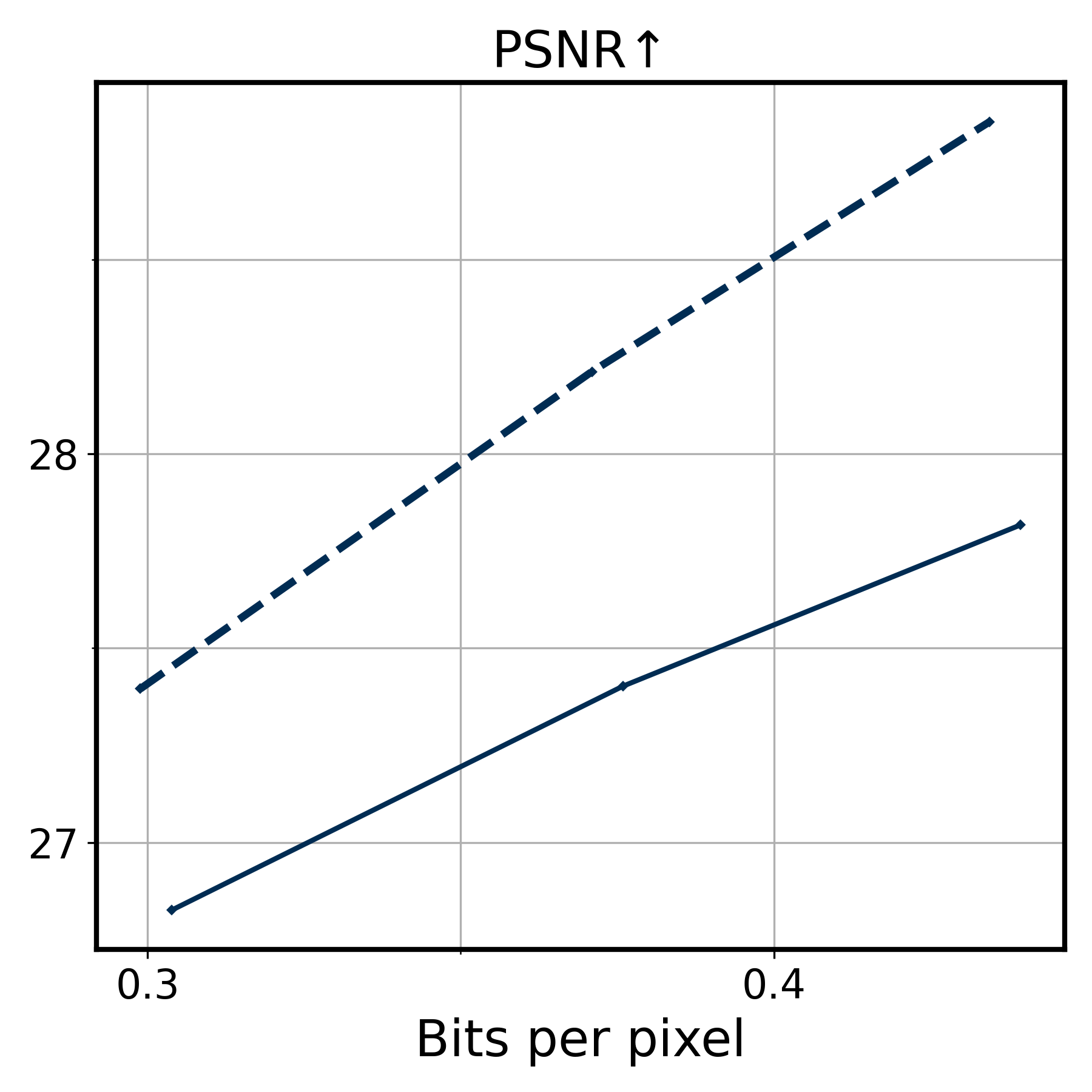}
    \end{minipage}
    \begin{minipage}{0.18\textwidth}
        \centering
        \includegraphics[width=\linewidth]{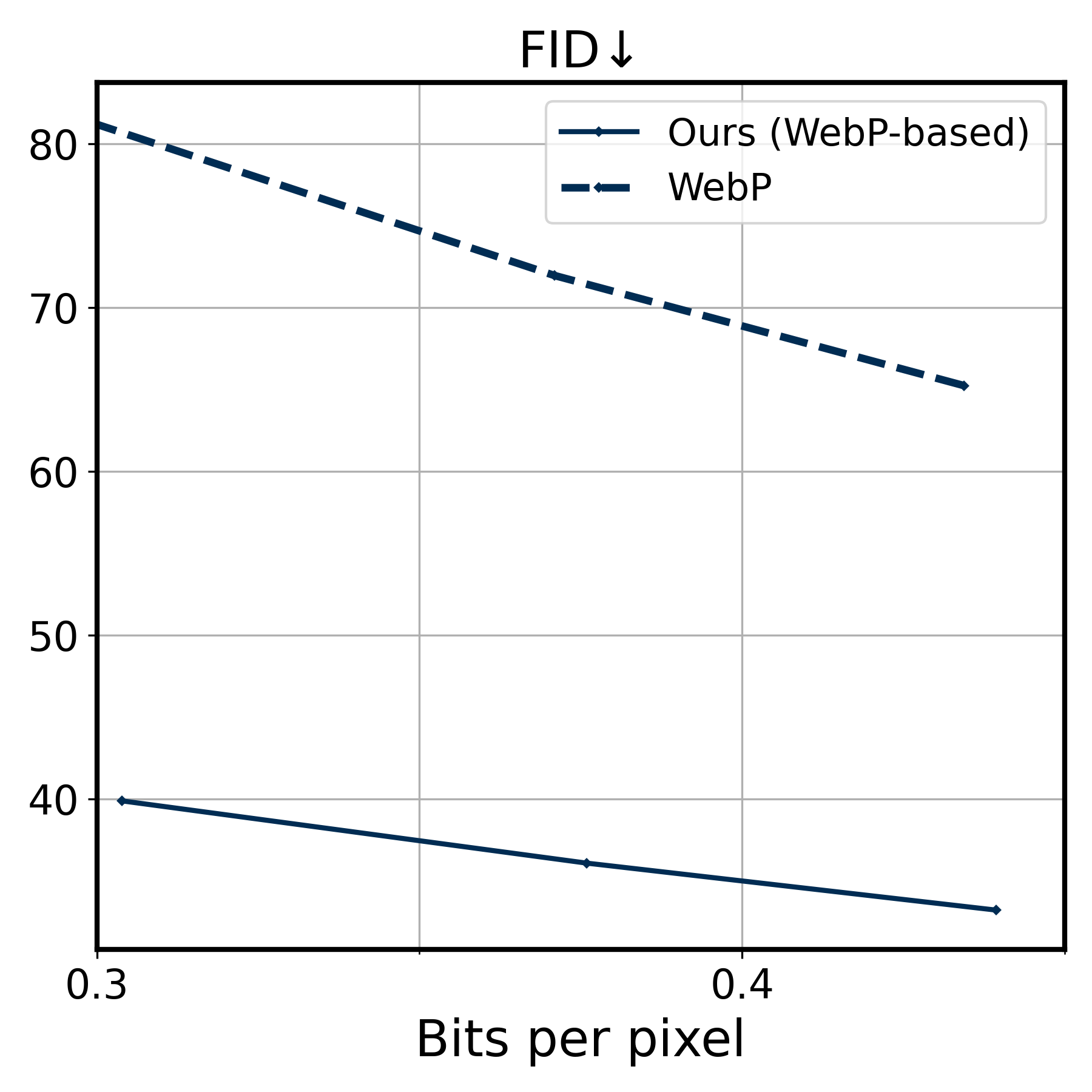}
    \end{minipage}
    \caption{Test our UniMIC on WebP. The significant improvement in FID shows our codec repository is extensible.}
    \label{fig:ablation_sp_webp}
\end{figure}

\myparagraph{Extensible codec repository.} 
Codecs are various in practical application, which requires that our codec repository should be flexible, allowing for the easy addition of any new codecs. Fig.~\ref{fig:ablation_sp_webp} illustrates this point with testing based on WebP.

\section{Limitation}
\label{sec: Limitation}
A major limitation of our UniMIC is the inference time. When using DDIM sampler with 50 steps, the time cost is about 14 seconds for a 512$\times$768 image on an NVIDIA RTX 3090 GPU. However, an amount of research focused on accelerating the diffusion process, including distillation~\cite{salimans2022progressive,meng2023distillation}, consistency models~\cite{song2023consistency,latentconsistency}, and faster samplers~\cite{dpm,dpm++}, promoting the completion of the diffusion process in just a few steps. These techniques are orthogonal to our approach and can be directly applied to accelerate our model.
\section{Conclusion}
\label{sec:conclusion}
In this paper, we propose UniMIC, the first universal multi-modality perceptual image compression framework, which establishes unified rate-distortion-perceptual (RDP) optimization for all image codecs in our compositional codec repository. To achieve universal multi-modality compression, we introduce multi-grained textual coding. This involves three levels of variable-length content prompt coding to accommodate different bitrate requirements and transfer content information to the decoding side for text-assisted decoding. We also design a compression prompt coding scheme to ensure the adaptability and applicability of our UniMIC across different basic codecs. For great RDP performance, we develop a universal perception compensator that leverages the cross-modality generation priors of diffusion models. This is achieved through the universal visual adapter, guided by textual information. 
{
    \small
    \bibliographystyle{ieeenat_fullname}
    \bibliography{main}
}
\clearpage
\setcounter{page}{1}
\maketitlesupplementary

\begin{figure*}[t]
    \centering
    \begin{minipage}{0.245\textwidth}
        \centering
        \includegraphics[width=\linewidth]{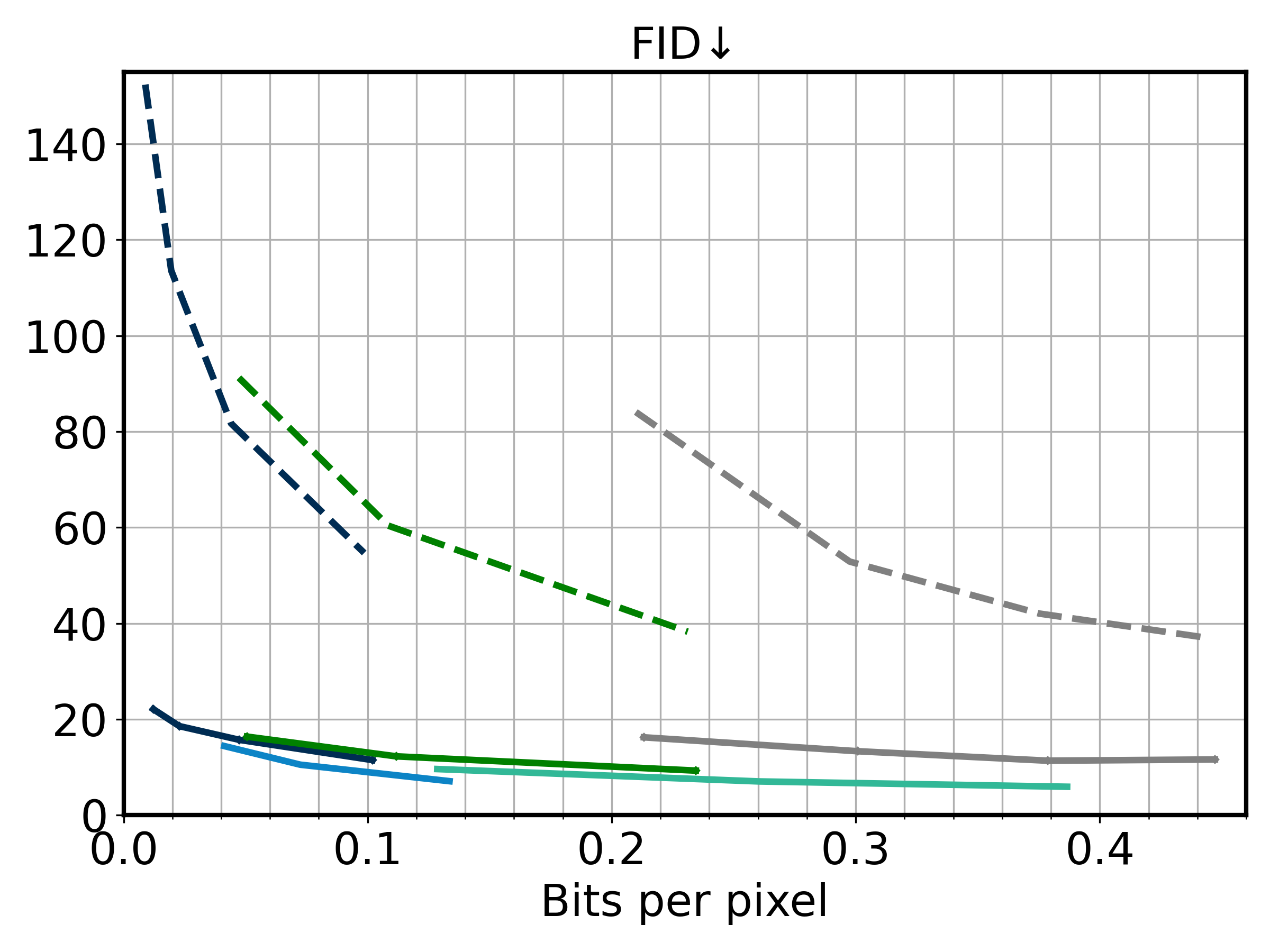}
    \end{minipage}
    \begin{minipage}{0.245\textwidth}
        \centering
        \includegraphics[width=\linewidth]{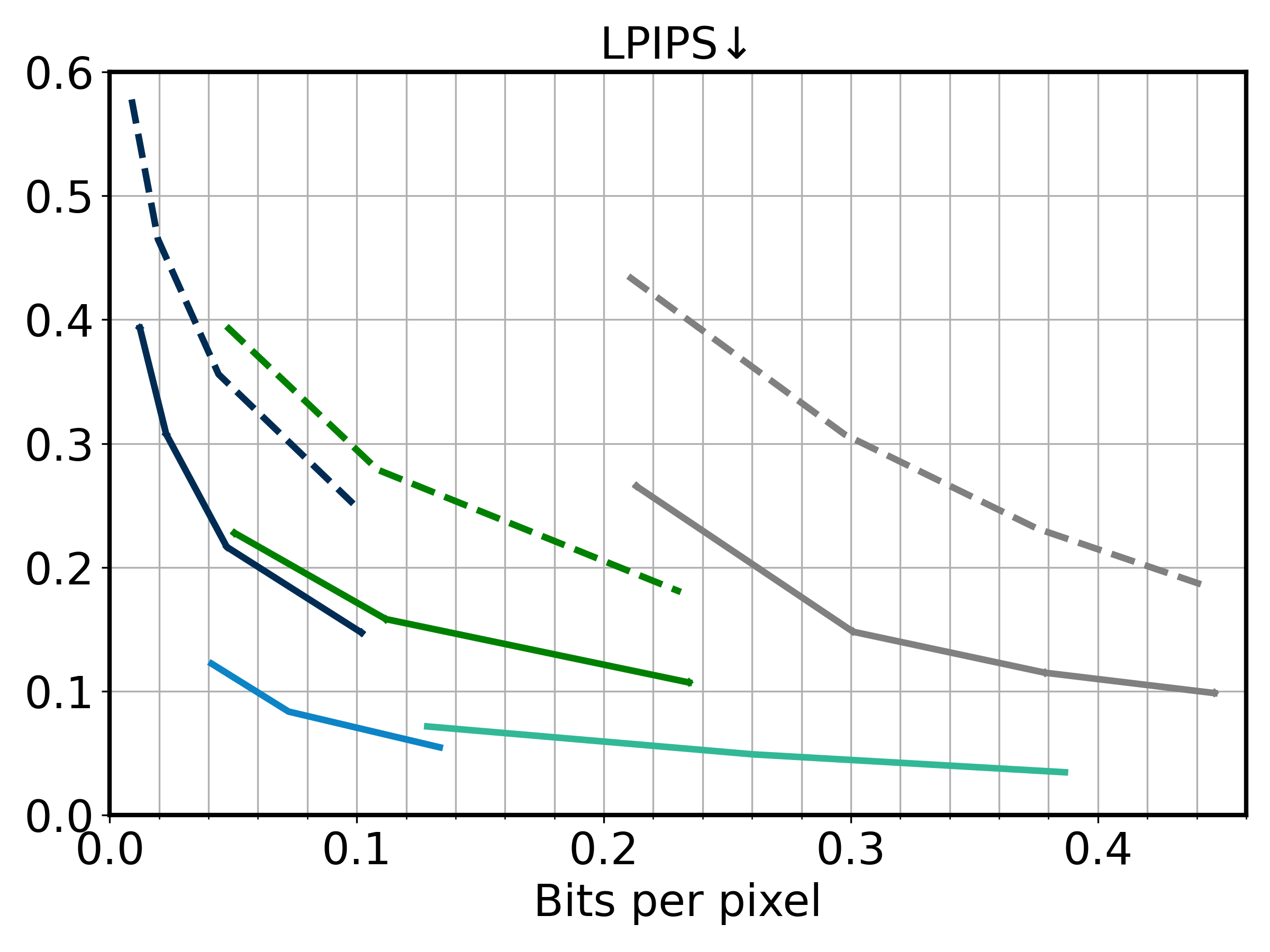}
    \end{minipage}
    \begin{minipage}{0.245\textwidth}
        \centering
        \includegraphics[width=\linewidth]{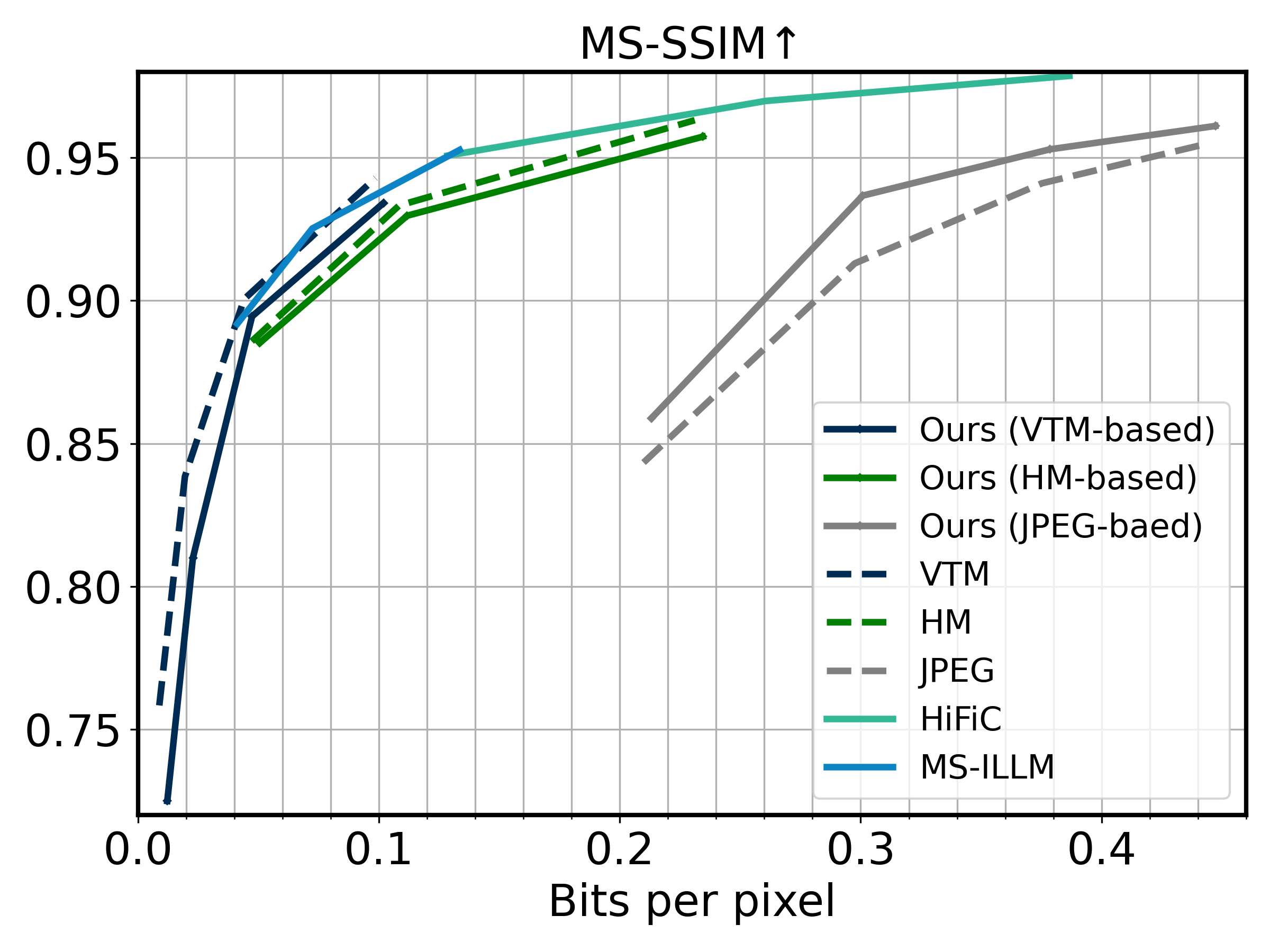}
    \end{minipage}
    \begin{minipage}{0.245\textwidth}
        \centering
        \includegraphics[width=\linewidth]{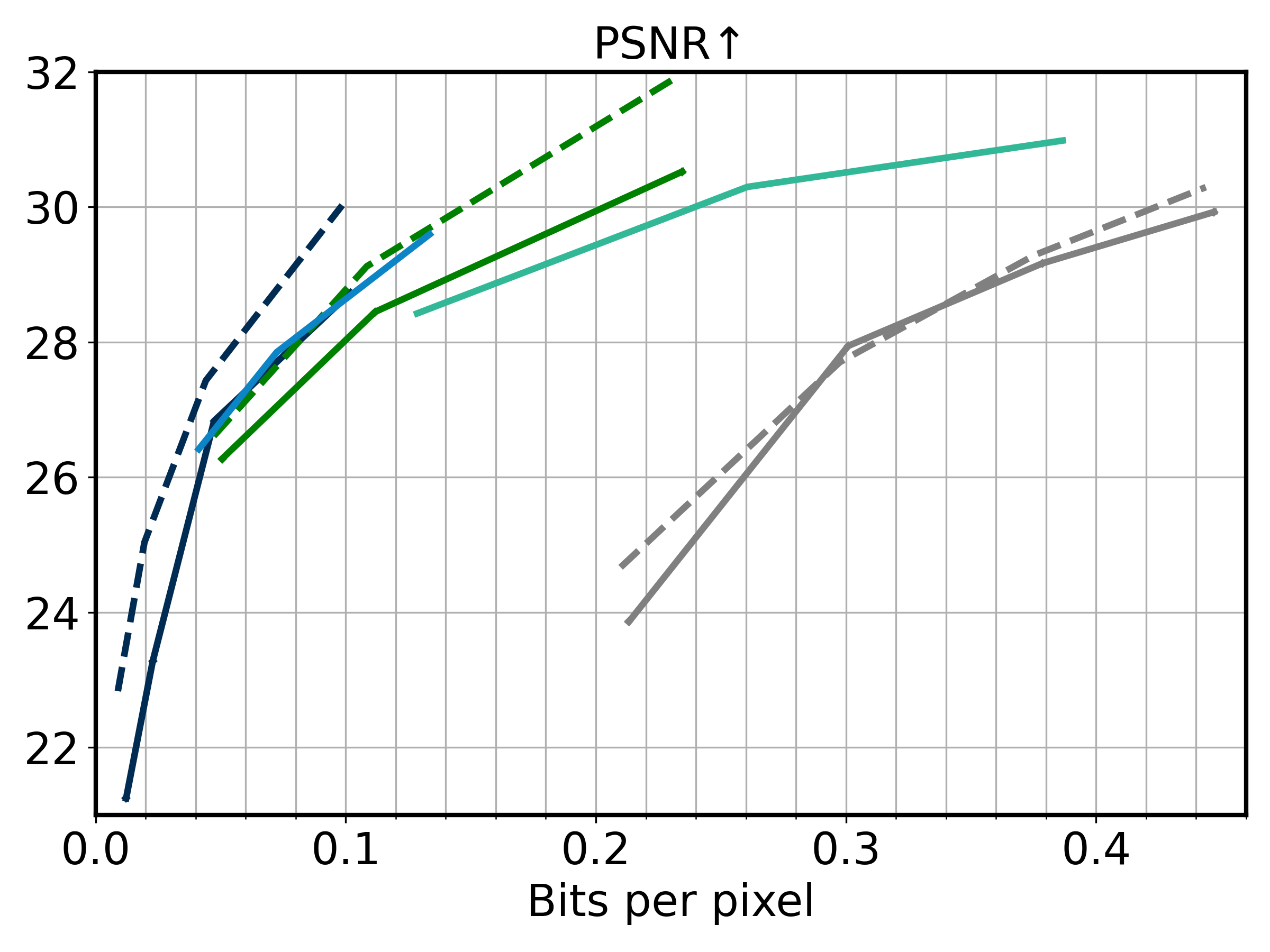}
    \end{minipage}
    \begin{minipage}{0.245\textwidth}
        \centering
        \includegraphics[width=\linewidth]{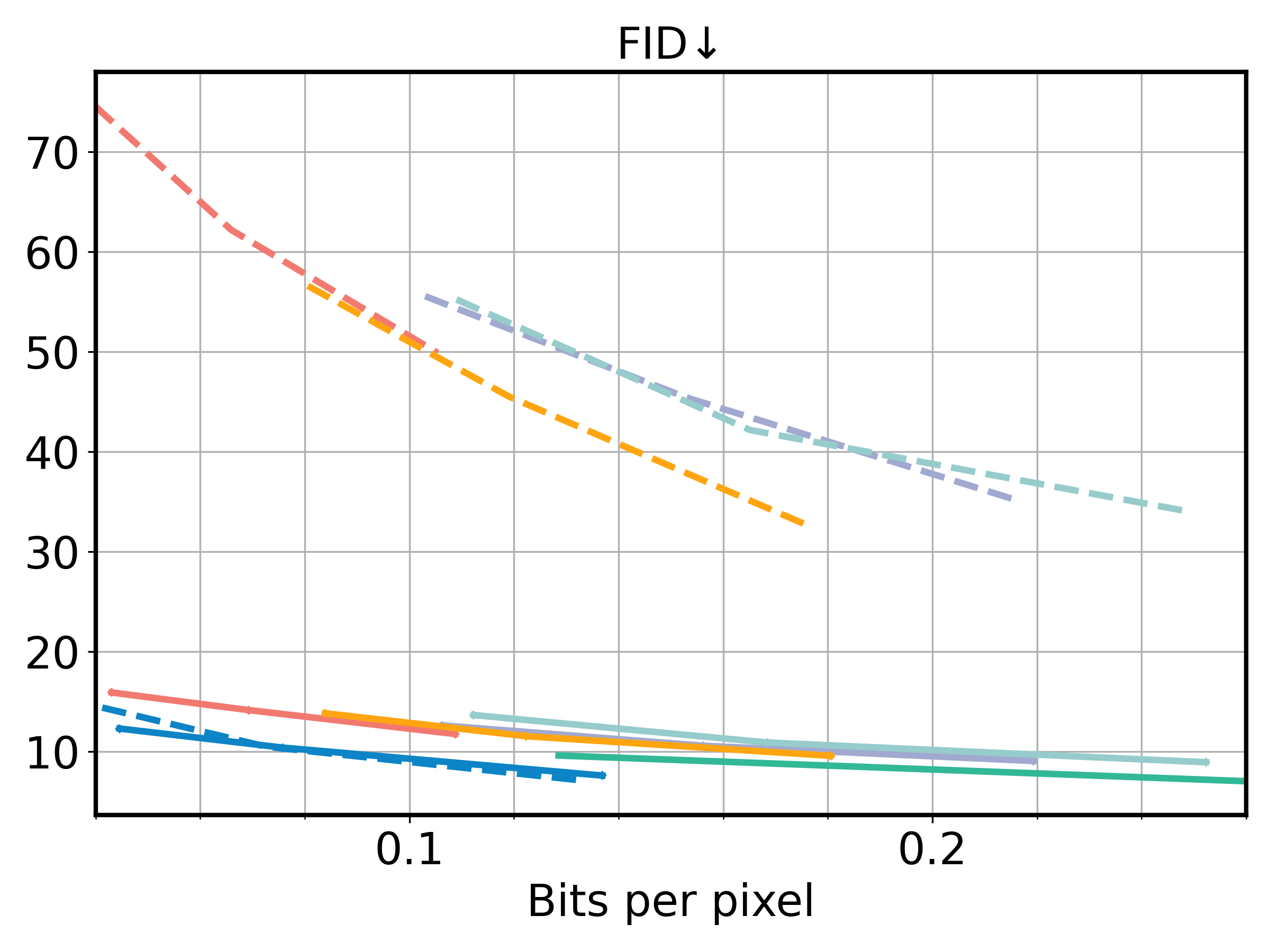}
    \end{minipage}
    \begin{minipage}{0.245\textwidth}
        \centering
        \includegraphics[width=\linewidth]{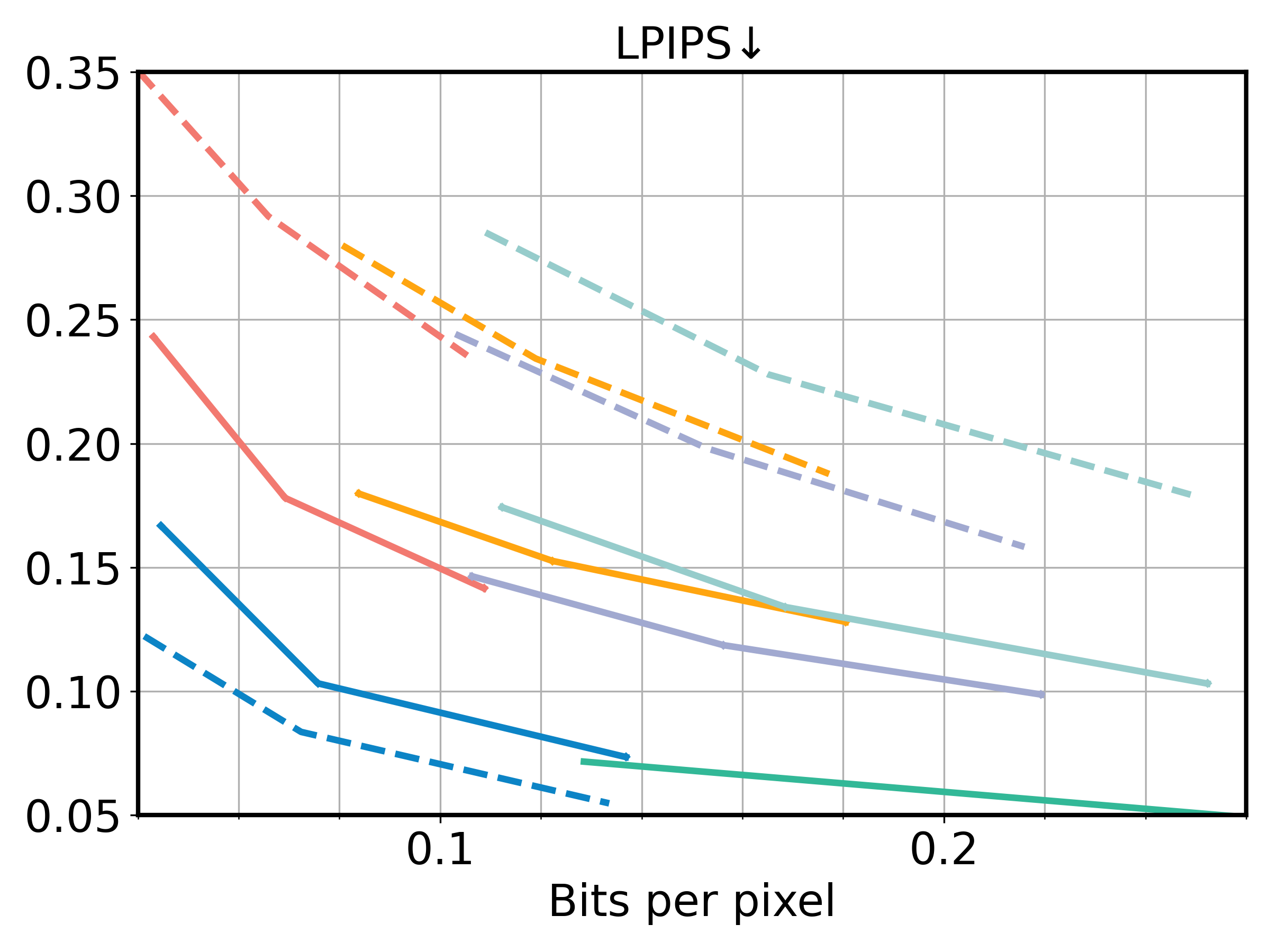}
    \end{minipage}
    \begin{minipage}{0.245\textwidth}
        \centering
        \includegraphics[width=\linewidth]{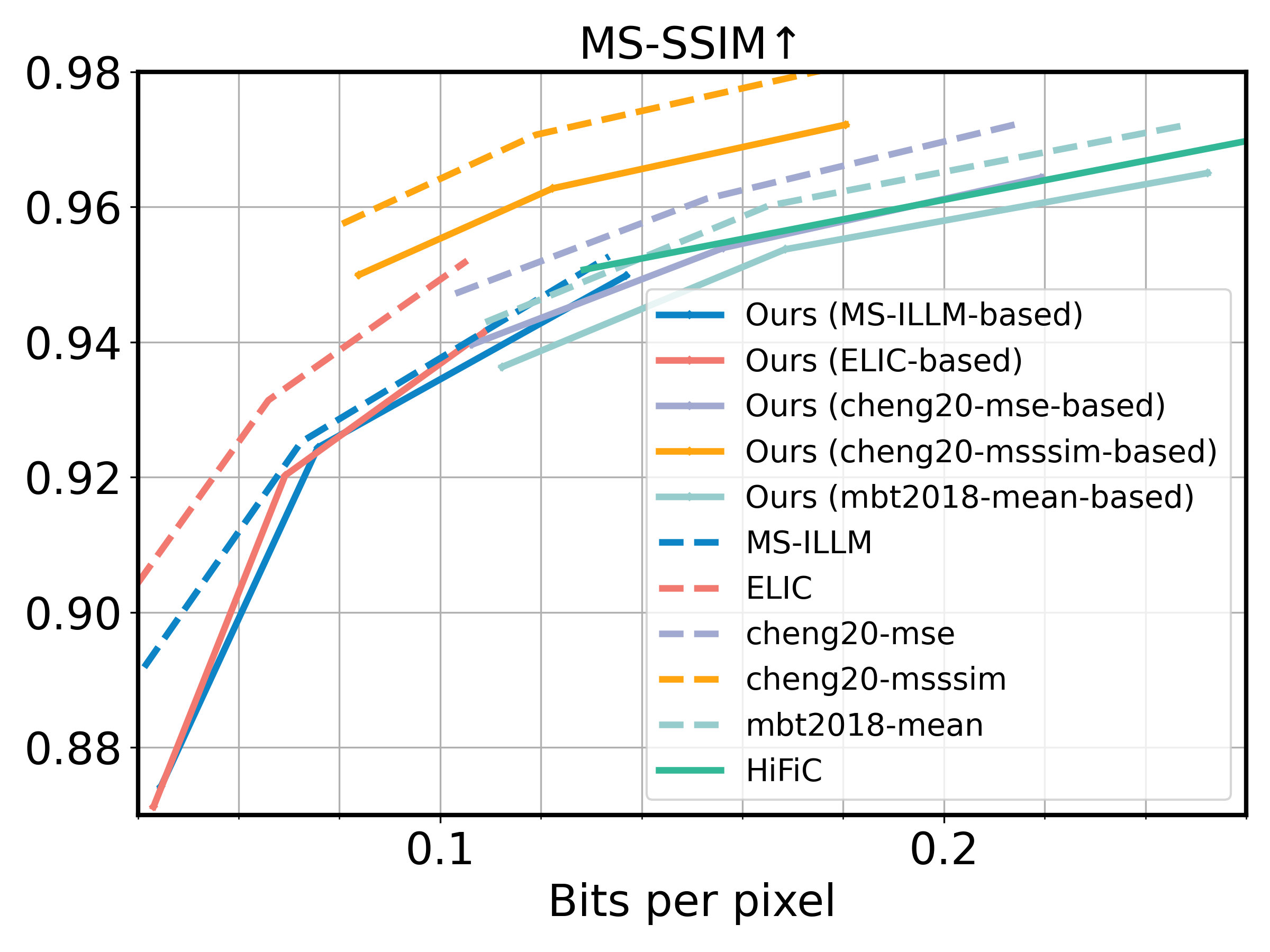}
    \end{minipage}
    \begin{minipage}{0.245\textwidth}
        \centering
        \includegraphics[width=\linewidth]{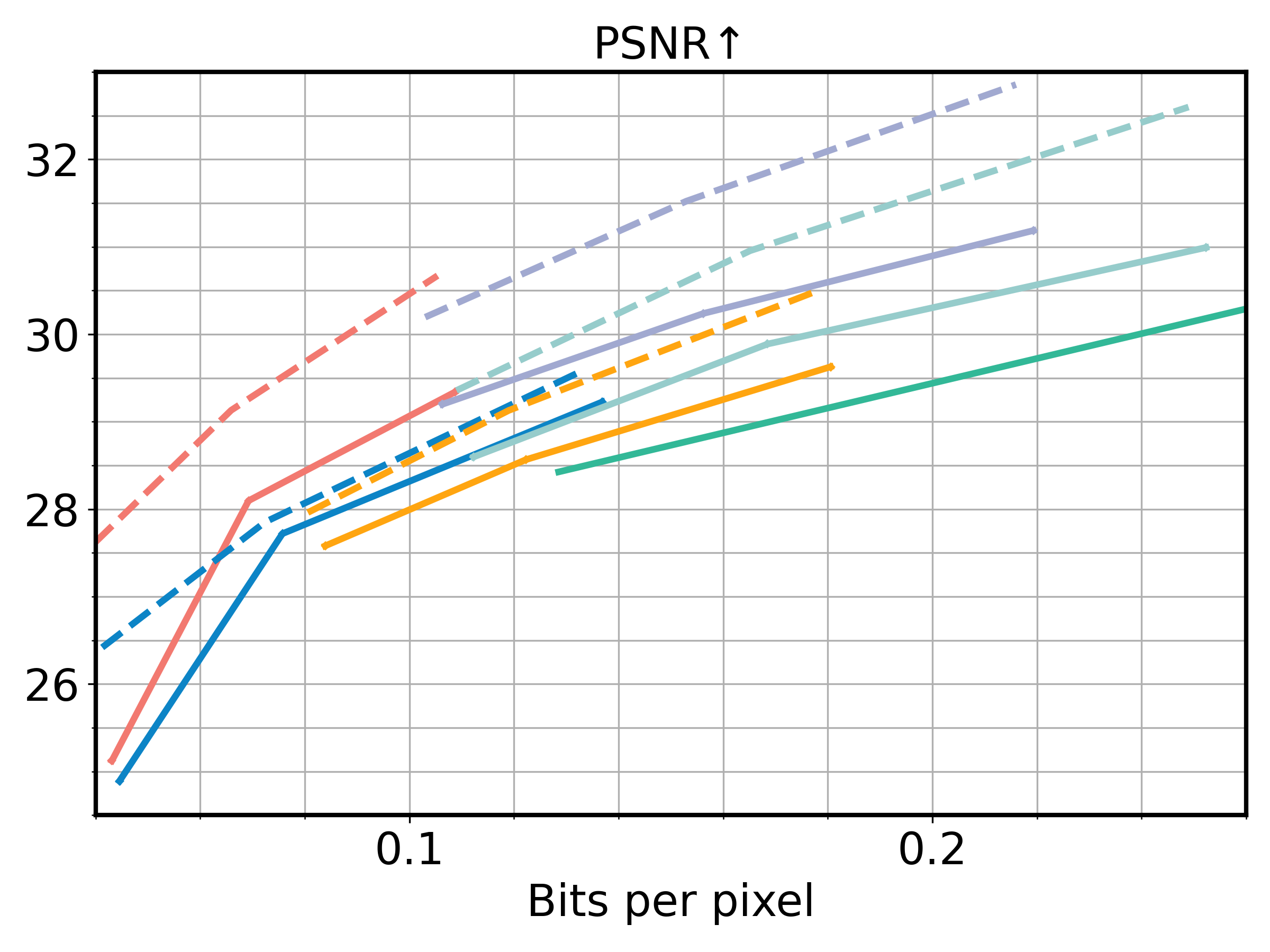}
    \end{minipage}
    \caption{Overall performance comparison between our method and state-of-the-art codecs on CLIC20 test split.}
    \label{fig:main_results_clic}
\end{figure*}

\begin{table*}[ht]
\resizebox{1.0\textwidth}{!}{
\begin{tabular}{cccccccccc}
\toprule
Method   & JPEG           & HM-intra 18.0  & VTM-intra 23.1 & mbt2018-mean   & cheng20-anchor-mse & cheng20-anchor-msssim & ELIC           & MS-ILLM       & Average        \\
\midrule
StableSR & -40.43 / \tcr{\textbf{-0.05}} & -53.07 / -0.99 & -91.13 / \tcr{\textbf{-1.09}} & -45.55 / -1.27 & \tcr{\textbf{-55.35}} / -1.80     & -51.26 / -0.66        & -70.10 / -1.38 & -1.75 / \tcr{\textbf{-0.44}} & -51.08 / -0.96 \\
SUPIR    & -32.01 / -3.70 & -38.27 / -3.70 & -56.53 / -1.61 & -29.30 / -5.26 & -36.38 / -5.72     & -41.70 / -4.13        & -43.00 / -3.33 & 3.48 / -2.48  & -34.21 / -3.74 \\
Ours   & \tcr{\textbf{-48.27}} / -0.19 & \tcr{\textbf{-55.48}} / \tcr{\textbf{-0.81}} & \tcr{\textbf{-96.91}} / -1.40 & \tcr{\textbf{-46.52}} / -1.09 & -54.82 / \tcr{\textbf{-1.37}}     & \tcr{\textbf{-51.60}} / \tcr{\textbf{-0.59}} & \tcr{\textbf{-72.08}} / \tcr{\textbf{-1.31}} & \tcr{\textbf{-3.71}} / -0.62 & \tcr{\textbf{-53.67}} / \tcr{\textbf{-0.92}} \\
\bottomrule
\end{tabular}}
\caption{Comparison between our method and other pose-processing methods. The evaluation metrics are presented as BD-FID(\%) (lower is better) / BD-PSNR(\%) (higher is better). The values are calculated between the results of each base codec and our UniMIC. \tcr{\textbf{Red}} color represents the best performance.}
\label{tab:post}
\end{table*}

In this document, we provide the supplementary material for the proposed universal multi-modality image compression framework, UniMIC. This includes the further experimental details, additional experimental results.


\section{Experiments}
\subsection{Further Experimental Details}
\myparagraph{Implementation Details.} 
The visual codec repository composes eight representative codecs in training: 1) traditional codecs: JPEG~\cite{wallace1992jpeg} (QF$\in\{5,10,15,20\}$), HM-intra 18.0~\cite{HEVC} (QP$\in\{47,42,37\}$), VTM-intra 23.1~\cite{VVC} (QP$\in\{57,52,47,42\}$). neural image codecs: mbt2018-mean~\cite{minnen2018joint} ($\lambda\in\{0.0018,0.0035,0.0067\}$), cheng20-anchor-mse~\cite{cheng2020learned} ($\lambda\in\{0.0018,0.0035,0.0067\}$), cheng20-anchor-msssim ($\lambda\in\{2.40,4.58,8.73\}$), ELIC~\cite{he2022elic} ($\lambda\in\{0.0004,0.0008,0.0016\}$), MS-ILLM~\cite{msillm} ($q\in\{1,2,3\}$), where mbt2018-mean is Mean\&Scale version in~\cite{minnen2018joint} and cheng20-anchor is the varient model from~\cite{cheng2020learned} which has no self-attention. cheng20-anchor-mse means MSE-optimized cheng20-anchor and cheng20-anchor-msssim is MS-SSIM-optimized one.
QF means the Quality Factor in JPEG (lower is quality worse) and QP is the Quantization Parameter in HM-intra and VTM-intra (higher is quality worse). For neural codecs, $\lambda$ means the Lagrange multiplier of rate-distortion trade-off (lower is quality worse), which is formalized as $L=R+\lambda D$, where $R$ denotes rate and $D$ denotes distortion. For MS-ILLM, $q$ represents the quality level (lower is quality worse).

For JPEG, we use the \texttt{torch-fidelity} repository \url{https://github.com/Queuecumber/torchjpeg}.
For HM-intra 18.0, we use the software from \url{https://vcgit.hhi.fraunhofer.de/jvet/HM/-/releases/HM-18.0}.
For VTM-intra 23.1, we use the official VTM software repository from \url{https://vcgit.hhi.fraunhofer.de/jvet/VVCSoftware_VTM/-/releases/VTM-23.1}.

For mbt2018-mean, cheng20-anchor-mse, and cheng20-anchor-msssim, we use the publicly released checkpoints in the \texttt{CompressAI} repository~\cite{begaint2020compressai}. 
For ELIC, we used the publicly released checkpoints in the ELiC ReImplementation repository \url{https://github.com/VincentChandelier/ELiC-ReImplemetation}. 
For MS-ILLM, we use the publicly released checkpoints in the official repository ~\cite{muckley2021neuralcompression}. 
For HiFiC, we use the code and pretrained checkpoints from \url{https://github.com/Justin-Tan/high-fidelity-generative-compression}.

\myparagraph{Metric Computation}
For the calculation of LPIPS and FID, we used the \texttt{TorchMetrics} package (available from \url{https://github.com/Lightning-AI/torchmetrics}) to calculate all metrics on our paper.

\subsection{More Results}
\myparagraph{Quantitative Results on CLIC20.}
Fig.\ref{fig:main_results_clic} presents a comparison between our method and state-of-the-art codecs on the CLIC20 test split\cite{CLIC2020}. Similar to Fig. 3 in the main text, our UniMIC achieves significant improvements in perceptual quality (measured by LPIPS and FID) across all eight basic codecs, demonstrating the universality of our model. 

\myparagraph{Comparison with post-processing methods.}
We provide the comparison with state-of-the-art post-processing methods, StableSR~\cite{wang2024exploiting} and SUPIR~\cite{supir}, in Tab.~\ref{tab:post}. 
StableSR is reproduced on our training set. However, it is worth noting that, as SUPIR does not provide the official training code, we directly utilize the pre-trained model offered by the code repositories of SUPIR for sampling purposes. In Tab.~\ref{tab:post}, our model achieves the best performance in both average objective and perceptual quality (last column), demonstrating its universality and strong adaptation across multiple image codecs.

\myparagraph{Trade-off Between Distortion and Perception.}
By varying the trade-off parameter $\beta$ (as mentioned in Section 4.2), we achieve a flexible distortion-perception trade-off. Figure 2 shows the rate-distortion curves for the VTM basic codec. We consider five values for $\beta$ in the study (0, 0.25, 0.5, 0.75, 1.0). The results indicate that larger $\beta$ leads to better perceptual quality but increased distortion. 
\begin{figure}[H]
    \centering
    \begin{minipage}{0.20\textwidth}
        \centering
    \includegraphics[width=\linewidth]{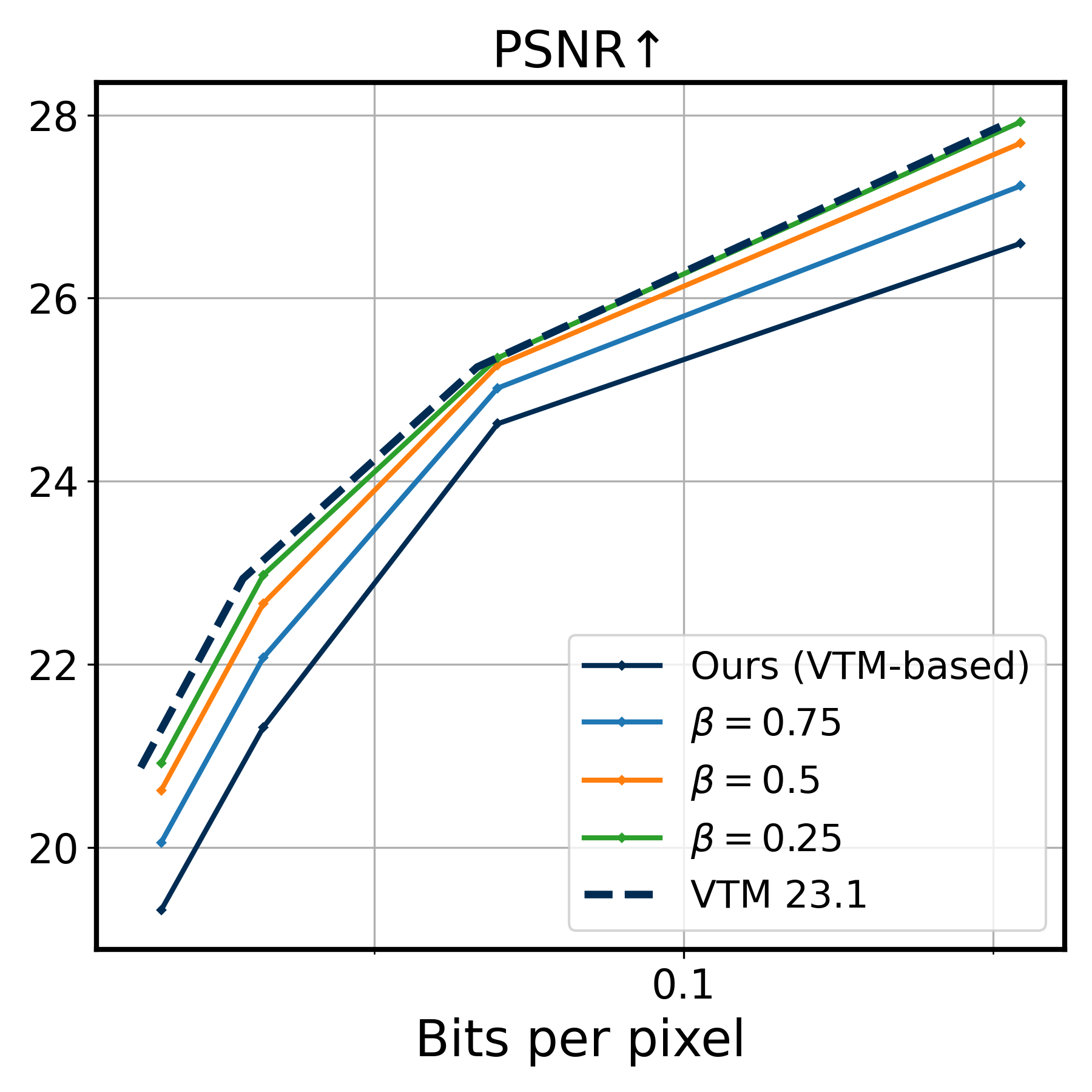}
    \end{minipage}
    \begin{minipage}{0.20\textwidth}
        \centering
        \includegraphics[width=\linewidth]{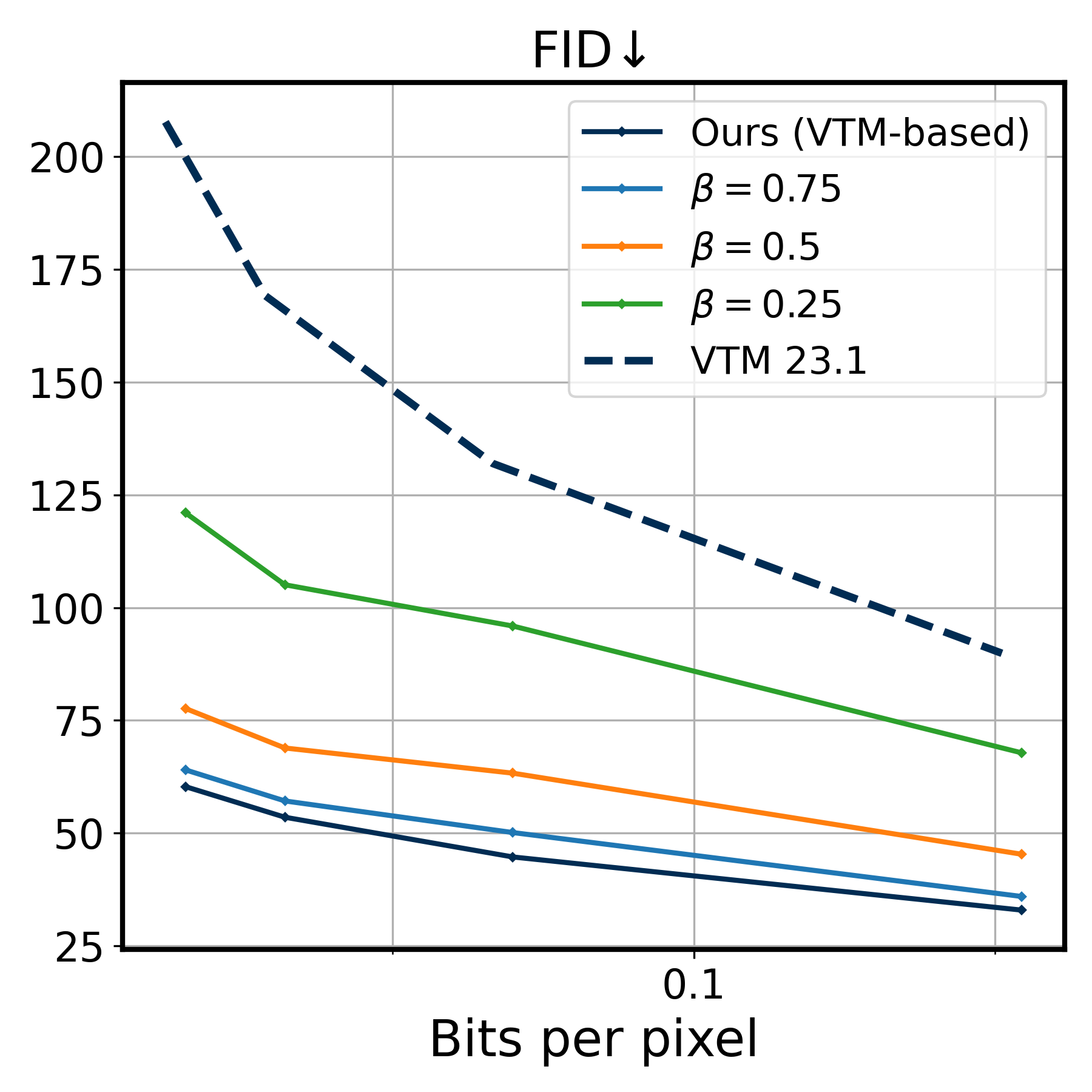}
    \end{minipage}
    \caption{Rate-distortion curves with different $\beta$ values.}
    \label{fig:inter}
\end{figure}

\begin{figure}[H]
    \centering
    \begin{minipage}{0.18\textwidth}
        \centering
    \includegraphics[width=\linewidth]{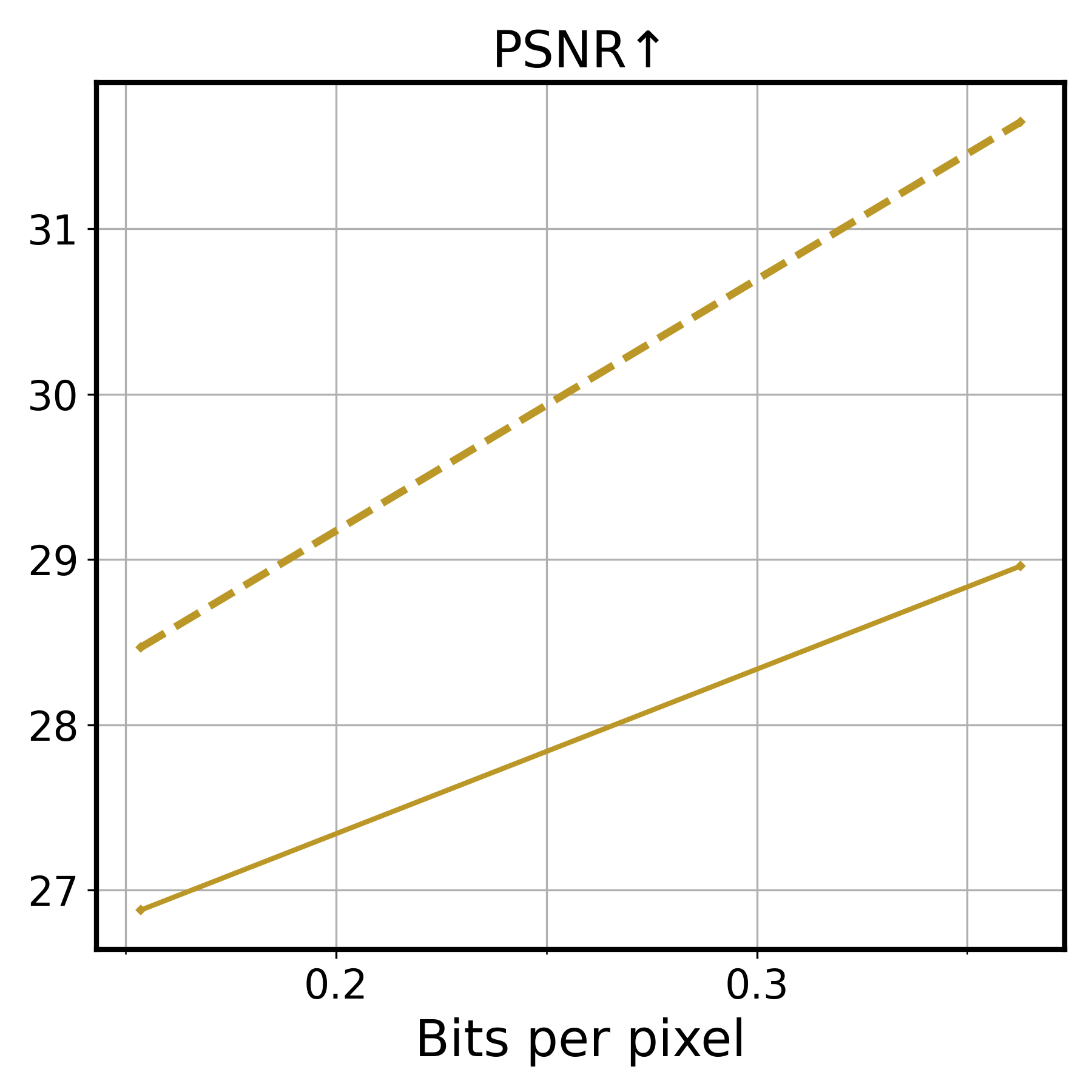}
    \end{minipage}
    \begin{minipage}{0.18\textwidth}
        \centering
        \includegraphics[width=\linewidth]{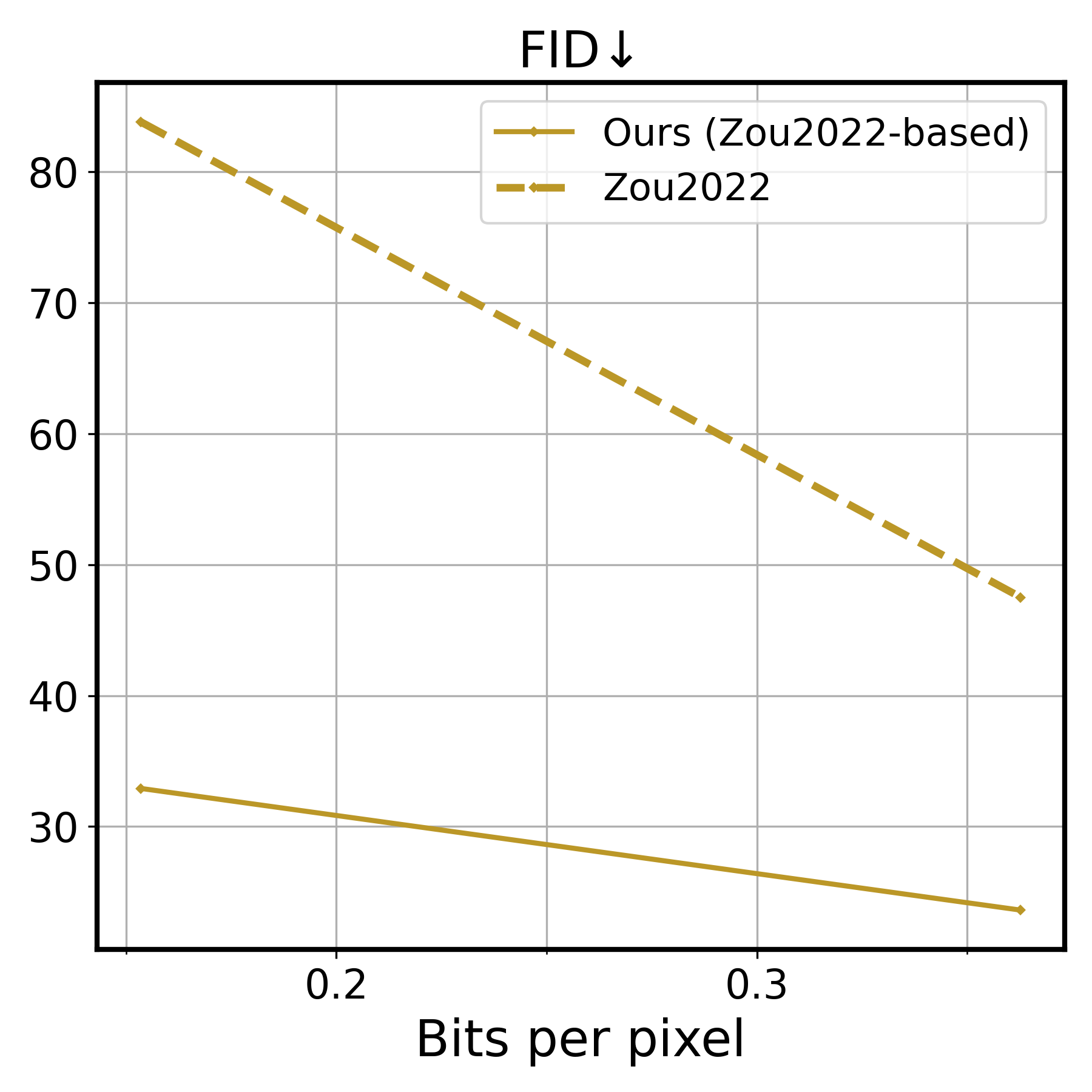}
    \end{minipage}
    \caption{Test our UniMIC on Zou2022~\cite{zou2022devil}. The significant improvement in FID shows our codec repository is extensible.}
    \label{fig:ablation_sp_zou2022}
\end{figure}

\myparagraph{Extensible codec repository.}
In the paper, we validated our extensible codec repository using a traditional image codec, WebP. Here, we provide additional results using a neural codec, Zou2022~\cite{zou2022devil}, which was also not part of the training set. Fig~\ref{fig:ablation_sp_zou2022} illustrates that our UniMIC significantly improves the FID value based on Zou2022, demonstrating the extensibility of our codec repository.


\end{document}